\documentclass[aps,longbibliography,
  twocolumn,amsmath,
  amssymb,nofootinbib, superscriptaddress]{revtex4-1}

\usepackage{hyperref}
\usepackage{graphicx}      
\usepackage{bm}            
\usepackage{xcolor}

\newcommand{\G}{G}

\begin{document}
\title{On the occurrence of fast neutrino flavor conversions\\
  in multidimensional supernova models}

\newcommand*{\APC}{Astro-Particule et Cosmologie (APC), CNRS UMR 7164,
Universit\'e Denis Diderot, 75205 Paris Cedex 13, France}
\newcommand*{\UNM}{Department of Physics \& Astronomy, University of New
 Mexico, Albuquerque, New Mexico 87131, USA}
\newcommand*{\Numazu}{Numazu College of Technology, Ooka 3600, Numazu, Shizuoka 410-8501, Japan}
\newcommand*{\NAOJ}{National Astronomical Observatory of Japan, Osawa, Mitaka, Tokyo 181-8588, Japan}
\author{Sajad Abbar}
\affiliation{\APC}
\affiliation{\UNM}
\author{Huaiyu Duan}
\affiliation{\UNM}
\author{Kohsuke Sumiyoshi}
\affiliation{\Numazu}
\author{Tomoya Takiwaki}
\affiliation{\NAOJ}
\author{Maria Cristina Volpe}
\affiliation{\APC}

\begin{abstract}
  The dense neutrino medium in a core-collapse supernova or a neutron-star
  merger event can experience fast flavor conversions on
  time/distance scales that are much smaller than those of vacuum
  oscillations. It
  is believed that fast neutrino flavor transformation occurs in the
  region where the angular
  distributions of $\nu_e$ and $\bar\nu_e$ cross each
  other. We present the first study of this crossing phenomenon and
  the fast neutrino flavor conversions in multidimensional (multi-D)
  supernova models.
  We examine the neutrino distributions obtained by solving
  the Boltzmann transport equation for several fixed
  profiles which are representative snapshots taken from separate 2D
  and 3D supernova  simulations with an $11.2 M_\odot$ progenitor model.
  Our research shows that the spherically
  asymmetric patterns of the $\nu_e$ and $\bar\nu_e$ fluxes in multi-D
  models can assist the appearance of the crossing between the
  $\nu_e$ and $\bar\nu_e$ angular distributions. In the models
  that we have studied,
  there exist unstable neutrino oscillation modes
  in and beyond the neutrino decoupling region which have amplitude growth
  rates as large as an $e$-fold per nanosecond.
  This finding can have important consequences for the
  explosion mechanism, nucleosynthesis, and neutrino signals of
  core-collapse supernovae.
\end{abstract}

\maketitle


\section{Introduction}
At the exhaustion of its nuclear fuel, the stellar core of a massive
star collapses under its own gravity
into a neutron star or black hole, and the rest of the star either explodes as
a supernova
or falls back to become a part of the black hole. In a successful
explosion, the hot proto-neutron star (PNS) formed
at the center of the (core-collapse) supernova quickly cools down by emitting
$\sim 10^{58}$ neutrinos in all
flavors in just tens of seconds. Although the exact mechanism of the
supernova explosion remains elusive,
the neutrino reactions
\begin{align}
  \bar\nu_e + p \rightleftharpoons n + e^+
  \quad\text{and}\quad
  \nu_e + n \rightleftharpoons p + e^-
  \label{eq:reaction}
\end{align}
are known  to be important in the heating and cooling of the matter
deep inside the
supernova and can play an important role in exploding the star
\cite{Bethe:1984ux,Janka:2012wk, Burrows:2012ew}.
Supernova ejecta are also one of the few places in the universe where the
heavy elements can be abundantly produced.
An important factor that regulates the nucleosynthesis in the
supernova ejecta is its electron fraction $Y_e$ (defined
as the ratio of the net number density of the electrons to that of the
baryons or $n_e/n_\mathrm{b}$) \cite{Qian:1996xt} which in turn is
determined by the neutrino processes in Eq.~\eqref{eq:reaction}. Because these
processes involve only $\nu_e$ and $\bar\nu_e$, and because the
neutrinos in different flavors are emitted in different intensities
and energies in a supernova, the transformation or oscillations between
different neutrino flavors
($\nu_e\rightleftharpoons\nu_{\mu/\tau}$ and
$\bar\nu_e\rightleftharpoons\bar\nu_{\mu/\tau}$)
near or even inside the surface of the PNS can have a significant
impact on the nucleosynthesis in the supernova ejecta and the
supernova dynamics.
Understanding the flavor transformation of the neutrinos inside the
supernova  is also crucial to
the prediction of the neutrino signals from future galactic supernovae and
the neutrino diffuse background
 \cite{Gava:2009pj, Horiuchi:2008jz, Beacom:2010kk,
   Mirizzi:2015eza,Horiuchi:2017qja}.

Because of the coherent forward scattering by the
dense neutrino medium surrounding the PNS,
supernova neutrinos can experience collective flavor transformation
\cite{Pastor:2002we,duan:2006an, duan:2006jv, duan:2010bg,
  Chakraborty:2016yeg} which occurs much deeper than
  does the flavor conversion   induced
by matter alone through the Mikheyev-Smirnov-Wolfenstein (MSW)
mechanism \cite{Wolfenstein:1977ue,Wolfenstein:1979ni,Mikheev:1986gs}.
However, based on the stationary neutrino bulb model \cite{duan:2006an} in
which all neutrinos are
decoupled from matter at a single sharp
neutrino sphere, it was found that collective flavor
transformation is
suppressed during the accretion phase of the supernova
due to the large density of the ambient matter
\cite{EstebanPretel:2008ni,
  Sarikas:2011am, Chakraborty:2011nf, Zaizen:2018wfg} and/or even the
neutrinos themselves \cite{Duan:2010bf}.
Although, in the more realistic models that evolve
with time and does not have the spherical symmetry,
this suppression can be lifted  for
the collective modes that oscillate rapidly in space and time
\cite{Duan:2014gfa,Chakraborty:2015tfa,Abbar:2015mca,Abbar:2015fwa, Dasgupta:2015iia},
the physical conditions can change significantly before
these modes have a sufficient growth in amplitudes to engender significant
flavor conversions.

In a real supernova, the neutrinos of different flavors (and energies)
decouple from matter at different depths of the PNS and have
different angular distributions outside the PNS. As a result,
fast neutrino flavor conversions can be
 driven by the neutrino density $n_\nu$ which occur on a
 characteristic
length scale $l_\text{fast}\sim (\hbar c)^{-2}G_\mathrm{F}^{-1} n_\nu^{-1}$ with
$G_\mathrm{F}$ being the Fermi coupling constant of the weak interaction
\cite{Sawyer:2005jk, Sawyer:2015dsa,
 Chakraborty:2016lct, Izaguirre:2016gsx,  Wu:2017qpc,
  Capozzi:2017gqd,
  Dasgupta:2016dbv, Abbar:2017pkh, Abbar:2018beu}.
Such flavor transformation is rightly called fast because
$l_\text{fast}$ is much shorter than the typical wavelength of
vacuum oscillations ($\sim\mathcal{O}(1)$ km for a 10
MeV neutrino with the atmospheric neutrino mass splitting)
and even the  mean free path of the neutrino inside the PNS
\cite{Capozzi:2018clo}.
It is believed that fast neutrino flavor conversions occur only where the
angular distributions of $\nu_e$ and $\bar\nu_e$  cross each other
\cite{Dasgupta:2016dbv,Izaguirre:2016gsx}.
However, an earlier study of the
one-dimensional (1D) supernova models of the Garching group with the
Boltzmann neutrino transport did not reveal any scenario with such
crossed neutrino angular distributions or
fast neutrino flavor conversions \cite{Tamborra:2017ubu}.
It has been
speculated  \cite{Chakraborty:2016lct, Dasgupta:2016dbv} that the multi-D models exhibiting
the lepton-emission
self-sustained asymmetry (LESA) \cite{Tamborra:2014aua} may have
regions that foster fast flavor conversions.
But the neutrino transports in most of the existing
multi-D supernova simulations are implemented with
various approximations and do not provide detailed angular
distributions of the neutrinos.

Recently, the full information of the neutrino angular distributions
became available to multi-D supernova models
by directly solving the Boltzmann equation in three momentum dimensions
\cite{Sumiyoshi:2012za}.
This approach has been used to study the neutrino transport
in supernova cores \cite{Sumiyoshi:2014qua}
and its impact on the explosion dynamics
\cite{Nagakura:2017mnp}.

In this paper, we present the first survey of the neutrino angular
distributions in multi-D supernova models
to ascertain the physical
conditions under which
fast neutrino flavor conversions may occur.

\section{Neutrino transport and fast flavor conversions}
At each space-time point $(t,\mathbf{r})$, the flavor content of the
neutrino medium in the momentum mode $\mathbf{p}$ can be specified by its flavor
density matrix $\varrho_{\mathbf{p}}(t, \mathbf{r})$
\cite{Sigl:1992fn} which, for two flavors ($e$ and $x$) and in the
weak-interaction basis, is \cite{Banerjee:2011fj}
\begin{align}
  \varrho = \frac{f_{\nu_e} + f_{\nu_x}}{2}
  + \frac{f_{\nu_e} - f_{\nu_x}}{2} \begin{bmatrix}
    s & S \\ S^* & -s \end{bmatrix},
\end{align}
where $f_{\nu_e/\nu_x}$ are the initial occupation numbers in the
corresponding flavors, and the complex and real scalar fields $S$ and
$s$ describe the flavor coherence and the flavor conversion of the
neutrino, respectively. A similar expression exists for the flavor
density matrix $\bar\varrho$ of the antineutrino.
The flavor evolution of the neutrino medium is governed by the
Liouville-von Neumann equation
\cite{Sigl:1992fn,Strack:2005ux,Cardall:2007zw,Volpe:2013jgr, Vlasenko:2013fja}
\begin{equation}
i (\partial_t + \mathbf{v} \cdot \bm{\nabla})
\varrho_{\mathbf{p}} = \left[
  \frac{\mathsf{M}^2}{2E_\nu} + \frac{\lambda}{2}\sigma_3 +
  \mathsf{H}_{\nu \nu, \mathbf{p}} ,
  \varrho_{\mathbf{p}}\right]
+ \mathcal{C}[\varrho_\mathbf{p}],
\label{Eq:EOM}
\end{equation}
where we have adopted the natural units $c=\hbar=1$.
In the above equation, $E_\nu=|\mathbf{p}|$,
$\mathbf{v} = \mathbf{p}/E_\nu$,  and
$\mathsf{M}^2$ are the energy,
velocity, and mass-square matrix of the neutrino, respectively,
$\lambda=\sqrt2 G_\mathrm{F} n_e$ is the matter potential
\cite{Wolfenstein:1977ue,Mikheev:1986gs}, $\sigma_3$ is the
third Pauli matrix,
\begin{equation}
  \mathsf{H}_{\nu \nu, \mathbf{p}} = \sqrt2 G_{\mathrm{F}}
  \int\!  \frac{\mathrm{d}^3p'}{(2\pi)^3}
  ( 1- \mathbf{v} \cdot \mathbf{v}')
  (\varrho_{\mathbf{p}'} - \bar\varrho_{\mathbf{p}'}).
\end{equation}
is the neutrino potential stemming from the neutrino-neutrino forward scattering
\cite{Fuller:1987aa,Notzold:1988kx,Pantaleone:1992xh},
and $\mathcal{C}[\varrho_\mathbf{p}]$ denotes the collision, emission
and absorption of the neutrinos.

To study fast neutrino flavor conversions, we ignore the vacuum
oscillation Hamiltonian $\mathsf{M}^2/2E_\nu$ so that the flavor
transformation of the neutrinos becomes energy independent. We also ignore
the collision term $\mathcal{C}[\varrho_\mathbf{p}]$ and
assume that the physical conditions are homogeneous and stationary for
the distance and time scales of interest
except for the
small but rapidly varying flavor mixing amplitude
$S_\mathbf{v}(t,\mathbf{r})$. This approximation is valid before
significant flavor conversion has occurred so that $|S_\mathbf{v}|\ll1$ and
$s\approx 1$. In this scenario, it is useful to define the angular
distribution of the
electron lepton number (ELN) of the neutrino flux  as
\cite{Izaguirre:2016gsx}
\begin{equation}
  \G_\mathbf{v} =
  \sqrt2 G_{\mathrm{F}}
  \int_0^\infty \frac{E_\nu^2 \mathrm{d} E_\nu}{(2\pi)^3}
        [f_{\nu_e}(\mathbf{p})- f_{\bar\nu_e}(\mathbf{p})],
 \label{Eq:G}
\end{equation}
where we have assumed $f_{\nu_x}(\mathbf{p})= f_{\bar\nu_x}(\mathbf{p})$.
Keeping only the terms in Eq.~\eqref{Eq:EOM} of magnitude
$\mathcal{O}(|S_\mathbf{v}|)$ or
larger, one obtains
\cite{Banerjee:2011fj, Vaananen:2013qja, Izaguirre:2016gsx}
\begin{equation}
 i (\partial_t + \mathbf{v} \cdot \bm{\nabla}) S_\mathbf{v}
 = (\epsilon_0 + \mathbf{v} \cdot
 \boldsymbol{\epsilon} ) S_\mathbf{v}
 - \int\!\mathrm{d}\Gamma_{\mathbf{v}'} (1 - \mathbf{v} \cdot \mathbf{v}')
 \G_{\mathbf{v}'} S_{\mathbf{v}'},
 \label{Eq:linear}
 \end{equation}
 where $\mathrm{d}\Gamma_{\mathbf{v}'}$ is the differential solid angle in the
 direction of $\mathbf{v}'$,
 $\epsilon_0 = \lambda+\int\!\mathrm{d}\Gamma_{\mathbf{v}'} \G_{\mathbf{v}'}$,
 and $\bm{\epsilon} = \int\!\mathrm{d}\Gamma_{\mathbf{v}'}
 \G_{\mathbf{v}'} \mathbf{v}'$.

Collective flavor transformation is induced by the normal modes in the neutrino
medium which, in the linear regime (where $|S_\mathbf{v}|\ll1$), are of the form
 $ S_\mathbf{v}(t,\mathbf{r}) = Q_\mathbf{v}\,  e^{-i\Omega t +
    i\mathbf{K}\cdot\mathbf{r}},$
where $Q_\mathbf{v}$, $\Omega$ and $\mathbf{K}$ are constant with the
latter two being also independent of $\mathbf{v}$. The flavor mixing
amplitude $S_\mathbf{v}$ remains small unless for a real wave vector
$\mathbf{K}$ the corresponding frequency has a positive imaginary
component, i.e.\
$\Omega_\mathrm{i}=\mathrm{Im}(\Omega) >0$. In this case the wave
amplitude can grow exponentially on the time scale of $l_\text{fast}/c$.
Such unstable normal modes with fast growing amplitudes can exist when
the ELN distribution $G_\mathbf{v}$ crosses 0 at some angle(s)
\cite{Dasgupta:2016dbv,Izaguirre:2016gsx}.


%
%
\begin{figure*} [tbh!]
 \centering
\begin{center}
\includegraphics*[width=1.\textwidth]{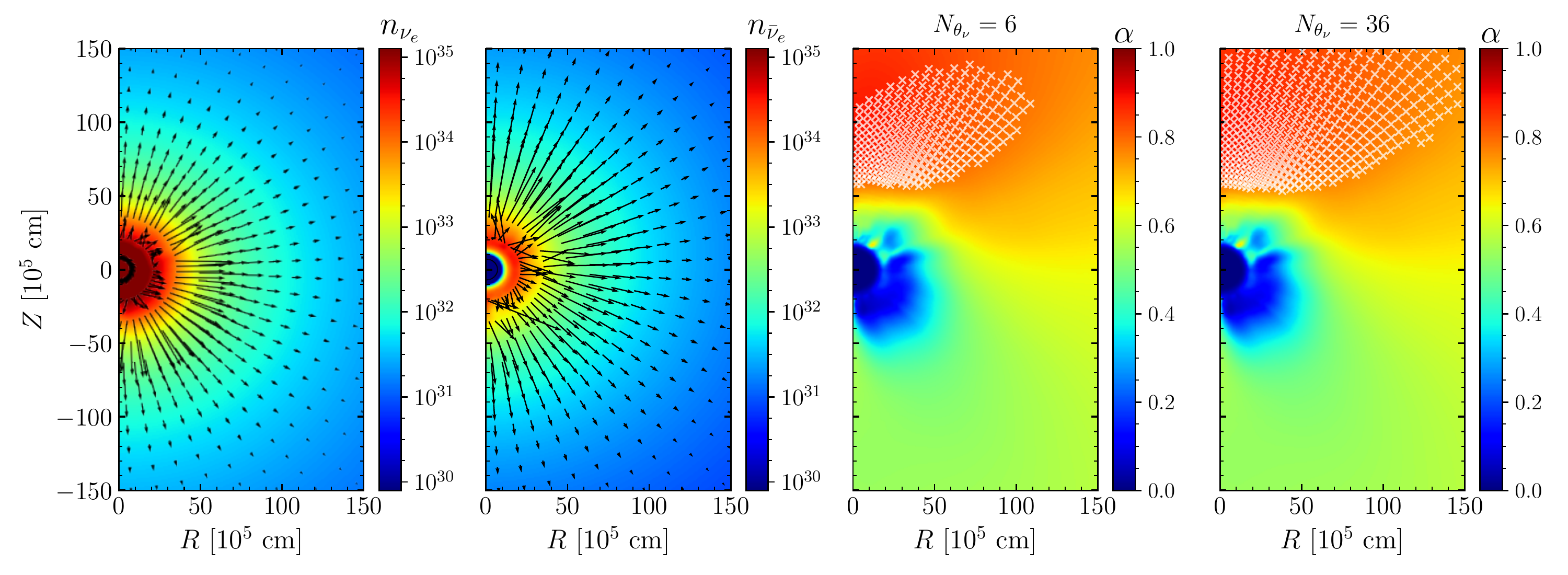}
\end{center}
\caption{The neutrino number densities $n_{\nu_e}$ and
  $n_{\bar\nu_e}$
  (left panels) and their ratio $\alpha=n_{\bar\nu_e}/n_{\nu_e}$
  (right panels) in the $t_\text{pb}=200$ ms snapshot of the 2D
  supernova model. The lengths and orientations of the arrows in the
  left panels indicate the magnitudes and directions of the average
  neutrino flux densities $\mathbf{j}_{\nu_e}$ and
  $\mathbf{j}_{\bar\nu_e}$ in the corresponding spatial zones. The white crosses
  in the right panels mark the zones where the ELN crossing
  occurs.
  There is almost no difference between angle averaged
  properties of the neutrinos
  solved in the low and high resolutions (with $N_{\theta_\nu}=6$ and 36,
  respectively)  except for the
  spatial extent over which the ELN crossing occurs.}
\label{fig:2D}
\end{figure*}

\section{ELN crossings in multi-D supernova models}
We study the angular distributions of the neutrinos which are obtained by
solving the Boltzmann transport equation (without any flavor
transformation) for several fixed supernova profiles \cite{Sumiyoshi:2014qua}.
These profiles were taken from the representative snapshots
at $t_\text{pb}=100$, 150 and 200 ms post the core bounce, respectively,
of a 2D and a 3D supernova simulations by using the
 Lattimer \& Swesty equation of state \cite{Lattimer:1991nc}
 with an approximate neutrino
transport and with an $11.2 M_\odot$ progenitor model \cite{Takiwaki:2011db,Takiwaki:2013cqa}.
The spatial resolutions of the Boltzmann calculations are $(256, 64, 1)$ and
$(256, 64, 32)$ for the
2D and 3D models, respectively, for radius up to $2613$ km from the
original simulations, where
$(N_r, N_\Theta, N_\Phi)$ are the numbers of spatial zones in the
spherical coordinates $(r, \Theta, \Phi)$. The momentum resolution
$(N_{E_\nu}, N_{\theta_\nu}, N_{\phi_\nu})$ of
the neutrino flux in each spatial zone is $(14, 6, 12)$, where
$\theta_\nu$ and $\phi_\nu$ are the zenith and azimuthal angles of the
neutrino velocity $\mathbf{v}$ with respect to the radial direction,
respectively.

Out of the three snapshots of the 2D model we find regions with ELN
crossings within and above the decoupling region
in the one at $t_\text{pb}=200$ ms. At this
time, the deformed shock has reached over 500 km and is poised to
explode the star with the help of a bipolar growth of the hydrodynamic
instabilities. Depending on their flavors and energies, neutrinos
decouple from matter at radius $\sim 50-70$ km
which can be viewed as the ``surface'' or neutrino sphere of the PNS.
In the first two panels of Fig.~\ref{fig:2D} we show the number
densities
$n_\nu = \int\!\frac{\mathrm{d}^3 p}{(2\pi)^3} f_\nu(\mathbf{p})$
and average flux densities
$\mathbf{j}_\nu = \int\!\frac{\mathrm{d}^3 p}{(2\pi)^3}
f_\nu(\mathbf{p}) \mathbf{v}$ in this snapshot
for $\nu=\nu_e$ and $\bar\nu_e$, respectively. Both $n_\nu$ and
$\mathbf{j}_\nu$ are mostly spherically symmetric in this snapshot
with $\mathbf{j}_\nu$ generally pointing in the radial direction
with some cases of non-radial fluxes.
Naively, one may think that
ELN crossings may occur below the neutrino sphere where  $\mathbf{j}_{\nu_e}$ and
$\mathbf{j}_{\bar\nu_e}$ can point in very different directions. This is not
the case, however, because $f_\nu(\mathbf{p})$ is highly isotropic
in this region, and we find no ELN crossing here.

In our study, ELN crossings usually begin to appear in the
region where the neutrinos begin to decouple from matter as shown
in the right panels of Fig.~\ref{fig:2D}.
Furthermore, we find
that, at the radii where ELN crossings do occur, they usually appear
in the angular zones with the $\bar\nu_e$-to-$\nu_e$ ratio
$\alpha=n_{\bar\nu_e}/n_{\nu_e}$ close to 1.
The correlation between $\alpha$ and ELN crossings is not really a
surprise. In the neutrino decoupling region, $f_\nu(\mathbf{p})$
becomes more and more peaked in the forward direction.
Because the PNS is rich in neutrons, $\bar\nu_e$'s decouple from matter
at smaller radii than $\nu_e$'s do and thus obtain a more forwardly peaked
distribution. However, this difference in $f_{\nu_e}(\mathbf{p})$ and
$f_{\bar\nu_e}(\mathbf{p})$  is usually not large enough to
result in an ELN crossing unless
$\bar\nu_e$ has a flux density very close to that of $\nu_e$. This is likely
the reason why no ELN crossing was found in a previous study of 1D
supernova models
\cite{Tamborra:2017ubu}. In the 2D model presented in
Fig.~\ref{fig:2D}, however, $\alpha$ can vary across angular zones at
the same radius which leads to ELN crossings in some regions.

\begin{figure*}[tbh!] 
\centering
\begin{center}
\includegraphics*[width=1.\textwidth]{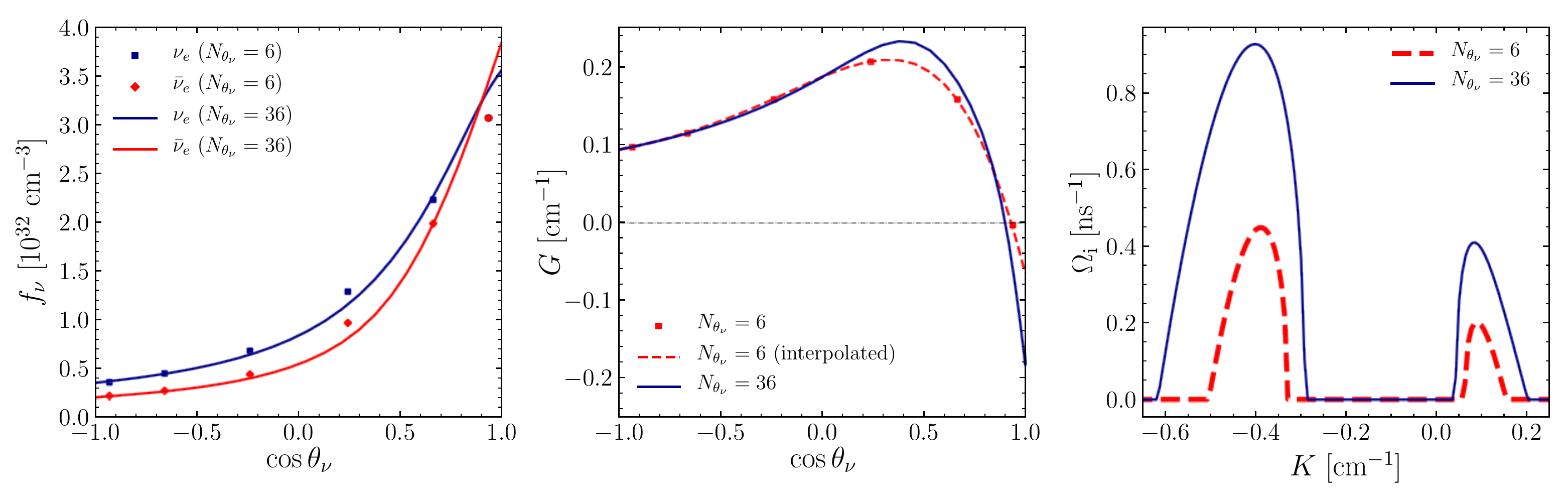}
\end{center}
\caption{The neutrino angular distributions $f_{\nu_e}(\theta_\nu)$
  and $f_{\bar\nu_e}(\theta_\nu)$ (left) and the ELN distribution
  $G(\theta_\nu)$ (middle) solved in two angular resolutions,
  $N_{\theta_\nu}=6$ and $36$, respectively, and the corresponding exponential
  growth rates $\Omega_\mathrm{i}$  as functions of the real wave number $K$ of
  the fast neutrino oscillation modes propagating in the
  radial direction. The results are for the spatial zone centered at
  $r=65.6$ km and $\cos\Theta=0.96$ in the 2D model shown
  in Fig.~\ref{fig:2D}.}
\label{fig:G}
\end{figure*}

\begin{figure*}[thb!] 
\centering
\begin{center}
\includegraphics*[width=1.\textwidth]{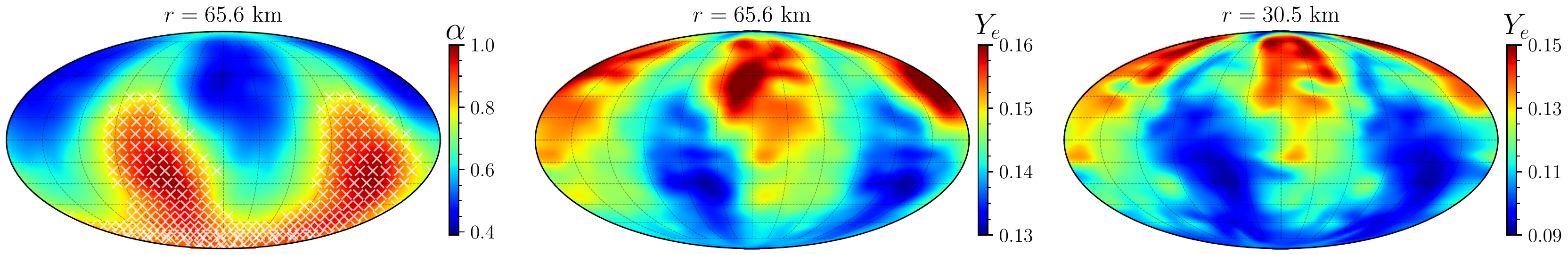}
\end{center}
\caption{The Mollweide projection of the $\bar\nu_e$-to-$\nu_e$
  density ratio $\alpha$ at $r=65.6$ km  (left) and
  the electron fractions $Y_e$ at $r=65.6$ km and $30.5$ km (middle and
  right), respectively, in the $t_\text{pb}=200$ ms snapshot of the 3D supernova
  model. The white crosses in the left panel mark the spatial zones
  where ELN crossings occur.}
\label{fig:3D}
\end{figure*}

To check the sensitivity of our
results on the angular resolution of the neutrino distributions, we
also solve the neutrino transport for the $t_\text{pb}=200$ ms
snapshot of the 2D model with $N_{\theta_\nu}=36$. We find
that, although
the angle averaged properties of the neutrinos such as $n_\nu$,
$\mathbf{j}_\nu$ and $\alpha$ are almost identical for the two
calculations, the
region with ELN crossings is much wider in the
high-resolution calculation than in the one with a lower
resolution (right panels of Fig.~\ref{fig:2D}). To find out the reason
why some spatial zones show ELN
crossings in the high-resolution calculation only, we compare the ELN
distributions $G_\mathbf{v}$ of these zones in the two calculations.
Because $f(\mathbf{p})$ are approximately axially
symmetric about the radial direction in our models, we
integrated them over $\phi_\nu$ and calculated
$f_\nu(\theta_\nu)=\iint\frac{E_\nu^2\mathrm{d}E_\nu \mathrm{d}\phi_\nu}{(2\pi)^3}
  f_\nu(\mathbf{p})$
and
$G(\theta_\nu)=\int_0^{2\pi}\mathrm{d}\phi_\nu G_\mathbf{v}$.
We plot $f_{\nu_e}(\theta_\nu)$, $f_{\bar\nu_e}(\theta_\nu)$ and
$G(\theta_\nu)$ for the spatial zone centered at $r=65.6$ km and
$\cos\Theta=0.96$ in the first two panels of Fig.~\ref{fig:G}. From this
figure one sees that the coarse angular resolution  is sufficient to
capture the overall shape of
$f_\nu(\theta_\nu)$, and it is more accurate in the
backward directions than in the forward
directions. However, because the ELN distribution $G(\theta_\nu)$ is
sensitive to the small difference between $f_{\nu_e}(\theta_\nu)$ and
$f_{\bar\nu_e}(\theta_\nu)$, and because it is likely to cross 0 near
the radial direction, a high angular resolution in the forward directions
is needed to accurately describe the crossing in the neutrino
 distributions. This issue becomes more severe at large radii
where $f_{\nu_e}(\theta_\nu)$ and $f_{\bar\nu_e}(\theta_\nu)$ are both
highly peaked in the forward direction, and they may cross each other
at an angle beyond the last $\theta_\nu$ bin.

We calculate the exponential growth rates $\Omega_\mathrm{i}$ of the
unstable fast
oscillation modes as functions of the real wave number $K$ for the
(interpolated)
ELN distributions plotted in Fig.~\ref{fig:G}, and the results are shown
in the right
panel in the same figure. We assume the
axial symmetry about the radial direction in calculating
$\Omega_\mathrm{i}$. We also assume that the
wave vector of the collective flavor oscillation wave is along the radial
direction. In this particular example,
$S_\mathbf{v}$ can grow by an  $e$-fold within $\sim 1$ nanosecond which
is indeed much faster than the typical changing rates of the physical conditions
inside the supernova.

We also find ELN crossings in the 3D model which
seem to be more common than in the 2D
model. This observation can be understood by noting that the spatial asymmetries in
the neutrino emission might evolve faster in the case of the 3D simulations \cite{Vartanyan:2018iah}.
They appear in all the three snapshots that we have studied. In the
$t_\text{pb}=200$ ms snapshot, ELN crossings begin to occur at $r=46$
km in the neutrino decoupling region.  We show $\alpha$ in this snapshot
at $r=65.6$ km in the left panel of Fig.~\ref{fig:3D} and mark with
crosses the spatial zones  where
ELN crossings occur. As in the 2D model, ELN crossings are more likely to
appear in the regions with  $\alpha$ close to 1.
The asymmetry pattern in the emissions of $\nu_e$ and $\bar\nu_e$ is
associated with a similar pattern in $Y_e$ which is present even well
below the neutrino sphere (middle and right panels of Fig.~\ref{fig:3D}).
The region with lower
$Y_e$ and more neutron-rich matter emits more $\bar\nu_e$'s and
absorbs more $\nu_e$'s through the reactions in
Eq.~\eqref{eq:reaction}. This pattern of $Y_e$ and $\alpha$ is likely
caused by the active matter convection inside the PNS as in the LESA
phenomenon \cite{Tamborra:2014aua}.

\begin{figure*}[tbh!] 
\centering
\begin{center}
\includegraphics*[width=1.\textwidth, trim=10 10 10 15,clip]{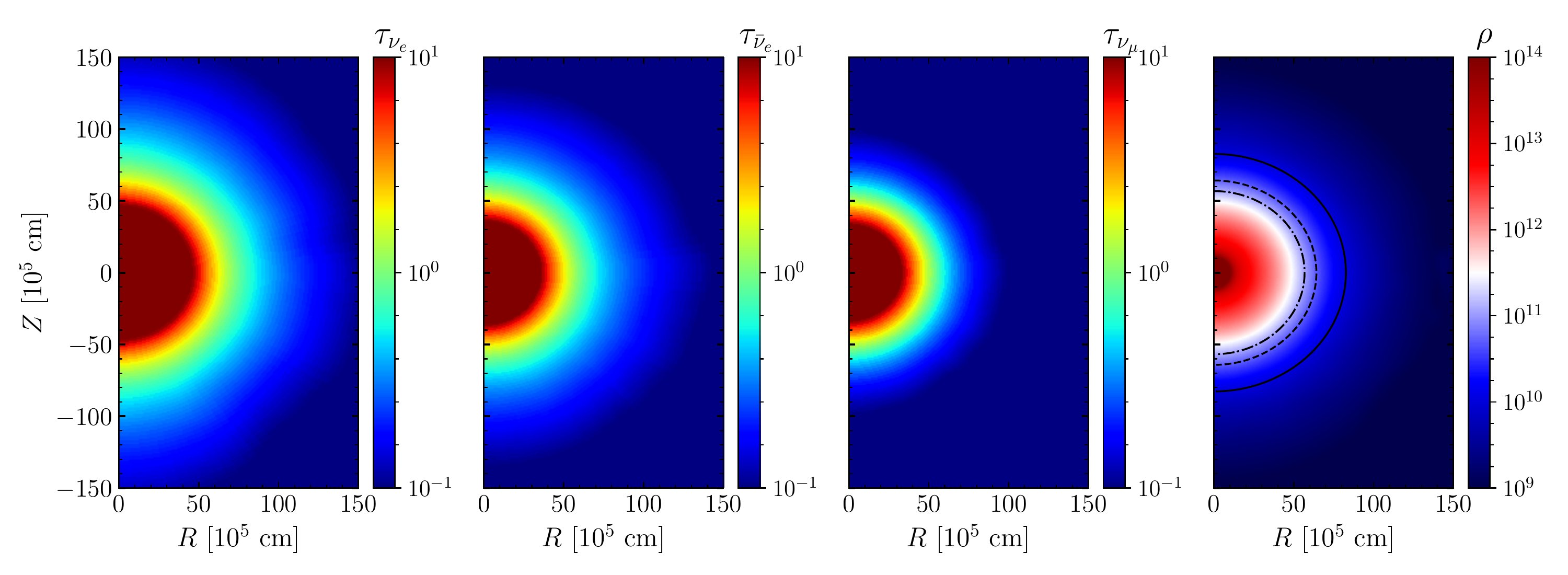}
\end{center}
\caption{The optical depths $\tau$ for the neutrinos of different flavors and
  with energy $E_{\nu} = 20.9$ MeV (the first three panels) and the
  matter density (the very right panel)
  in the $t_\text{pb}=200$ ms snapshot of the 2D supernova model shown
  in Fig.~\ref{fig:2D}.
  The solid, dashed and dot-dashed circles in the right panel indicate
  the location of the neutrinospheres of these neutrinos.
 }
\label{fig:optical}
\end{figure*}

Our study highlights the need of more multi-D supernova simulations
with the neutrino Boltzmann transport like those performed in
Ref.~\cite{Nagakura:2017mnp}. Our results show that the neutrino
distributions with even a very coarse angular resolution can be used
to identify the ELN crossings in the neutrino decoupling region,
although a better resolution in the
radial direction is needed to accurately describe the crossings
especially at large radii. Although we have found fast growing
neutrino oscillations in the multi-D models that we have studied, it
remains to be seen if the conditions for the ELN
crossing will be met in the more self-consistent multi-D supernova
simulations with accurate neutrino transport,
 and if the amplitudes of the fast neutrino oscillation  modes can
continue to grow into the nonlinear regime and cause significant
flavor conversions deep inside the supernova.
If fast neutrino flavor conversions are indeed found to occur near the
surface of the PNS, they will have profound implications for the
dynamics, nucleosynthesis, and neutrino signals of the supernova.

\section*{Acknowledgments}
We thank J.\ Martin and C.\ Yi for valuable discussions.
This work is supported by the US DOE EPSCoR grant DE-SC0008142 and the NP
grant DE-SC0017803 at UNM for S.~A.\ and H.~D.,
``Gravitation et physique
fondamentale’' (GPHYS) of the Observatoire de Paris for S.~A.\ and
M.~C.~V., and
JSPS and MEXT KAKENHI Grant Numbers
(JP15K05093, JP17H01130, JP17K14306,  18H01212)
and (JP26104006, 17H05206, JP17H06357, JP17H06364, JP17H06365)
for K.~S.\ and T.~T., resepctively.
K.~S.\ and T.~T.\ acknowledge the support on the computing resources
at NAOJ, KEK, JLDG, YITP, RCNP, UT
as well as Post-K and K-computer of the RIKEN AICS.

\appendix
\section{The optical depths}
In this appendix, we provide the information of the
optical depths in the $t_\text{pb}=200$ ms snapshot of the 2D
supernova model shown in Fig.~\ref{fig:2D}. The optical depth of a neutrino is
defined as
\begin{equation}
\tau(s) = \int_{s}^{\infty} \chi(s') \mathrm{d}s',
\end{equation}
where $\chi(s)$ is the neutrino opacity, and the integration is
performed along the radial coordinate for the neutrinos in the forward
angle bin.
In the first three panels of  Fig.~\ref{fig:optical},
we show the optical depths of the neutrinos of different flavors and
all with
energy $E_\nu = 20.9$ MeV. The very right panel of the figure shows the matter
density in this snapshot and  the location of the neutrinospheres
(where $\tau =2/3$) for these neutrinos.
The average energies of the neutrinos are in the range of $10-25$ MeV
in this snapshot.

\bibliography{fast_modes}

\begin{thebibliography}{56}%
\makeatletter
\providecommand \@ifxundefined [1]{%
 \@ifx{#1\undefined}
}%
\providecommand \@ifnum [1]{%
 \ifnum #1\expandafter \@firstoftwo
 \else \expandafter \@secondoftwo
 \fi
}%
\providecommand \@ifx [1]{%
 \ifx #1\expandafter \@firstoftwo
 \else \expandafter \@secondoftwo
 \fi
}%
\providecommand \natexlab [1]{#1}%
\providecommand \enquote  [1]{``#1''}%
\providecommand \bibnamefont  [1]{#1}%
\providecommand \bibfnamefont [1]{#1}%
\providecommand \citenamefont [1]{#1}%
\providecommand \href@noop [0]{\@secondoftwo}%
\providecommand \href [0]{\begingroup \@sanitize@url \@href}%
\providecommand \@href[1]{\@@startlink{#1}\@@href}%
\providecommand \@@href[1]{\endgroup#1\@@endlink}%
\providecommand \@sanitize@url [0]{\catcode `\\12\catcode `\$12\catcode
  `\&12\catcode `\#12\catcode `\^12\catcode `\_12\catcode `\%12\relax}%
\providecommand \@@startlink[1]{}%
\providecommand \@@endlink[0]{}%
\providecommand \url  [0]{\begingroup\@sanitize@url \@url }%
\providecommand \@url [1]{\endgroup\@href {#1}{\urlprefix }}%
\providecommand \urlprefix  [0]{URL }%
\providecommand \Eprint [0]{\href }%
\providecommand \doibase [0]{http://dx.doi.org/}%
\providecommand \selectlanguage [0]{\@gobble}%
\providecommand \bibinfo  [0]{\@secondoftwo}%
\providecommand \bibfield  [0]{\@secondoftwo}%
\providecommand \translation [1]{[#1]}%
\providecommand \BibitemOpen [0]{}%
\providecommand \bibitemStop [0]{}%
\providecommand \bibitemNoStop [0]{.\EOS\space}%
\providecommand \EOS [0]{\spacefactor3000\relax}%
\providecommand \BibitemShut  [1]{\csname bibitem#1\endcsname}%
\let\auto@bib@innerbib\@empty
\bibitem [{\citenamefont {Bethe}\ and\ \citenamefont
  {Wilson}(1985)}]{Bethe:1984ux}%
  \BibitemOpen
  \bibfield  {author} {\bibinfo {author} {\bibfnamefont {Hans~A.}\ \bibnamefont
  {Bethe}}\ and\ \bibinfo {author} {\bibfnamefont {James~R.}\ \bibnamefont
  {Wilson}},\ }\bibfield  {title} {\enquote {\bibinfo {title} {{Revival of a
  stalled supernova shock by neutrino heating}},}\ }\href {\doibase
  10.1086/163343} {\bibfield  {journal} {\bibinfo  {journal} {Astrophys. J.}\
  }\textbf {\bibinfo {volume} {295}},\ \bibinfo {pages} {14--23} (\bibinfo
  {year} {1985})}\BibitemShut {NoStop}%
\bibitem [{\citenamefont {Janka}(2012)}]{Janka:2012wk}%
  \BibitemOpen
  \bibfield  {author} {\bibinfo {author} {\bibfnamefont {Hans-Thomas}\
  \bibnamefont {Janka}},\ }\bibfield  {title} {\enquote {\bibinfo {title}
  {{Explosion Mechanisms of Core-Collapse Supernovae}},}\ }\href {\doibase
  10.1146/annurev-nucl-102711-094901} {\bibfield  {journal} {\bibinfo
  {journal} {Ann. Rev. Nucl. Part. Sci.}\ }\textbf {\bibinfo {volume} {62}},\
  \bibinfo {pages} {407--451} (\bibinfo {year} {2012})},\ \Eprint
  {http://arxiv.org/abs/1206.2503} {arXiv:1206.2503 [astro-ph.SR]} \BibitemShut
  {NoStop}%
\bibitem [{\citenamefont {Burrows}(2013)}]{Burrows:2012ew}%
  \BibitemOpen
  \bibfield  {author} {\bibinfo {author} {\bibfnamefont {Adam}\ \bibnamefont
  {Burrows}},\ }\bibfield  {title} {\enquote {\bibinfo {title} {{Colloquium:
  Perspectives on core-collapse supernova theory}},}\ }\href {\doibase
  10.1103/RevModPhys.85.245} {\bibfield  {journal} {\bibinfo  {journal} {Rev.
  Mod. Phys.}\ }\textbf {\bibinfo {volume} {85}},\ \bibinfo {pages} {245}
  (\bibinfo {year} {2013})},\ \Eprint {http://arxiv.org/abs/1210.4921}
  {arXiv:1210.4921 [astro-ph.SR]} \BibitemShut {NoStop}%
\bibitem [{\citenamefont {Qian}\ and\ \citenamefont
  {Woosley}(1996)}]{Qian:1996xt}%
  \BibitemOpen
  \bibfield  {author} {\bibinfo {author} {\bibfnamefont {Y.~Z.}\ \bibnamefont
  {Qian}}\ and\ \bibinfo {author} {\bibfnamefont {S.~E.}\ \bibnamefont
  {Woosley}},\ }\bibfield  {title} {\enquote {\bibinfo {title}
  {{Nucleosynthesis in neutrino driven winds: 1. The Physical conditions}},}\
  }\href {\doibase 10.1086/177973} {\bibfield  {journal} {\bibinfo  {journal}
  {Astrophys. J.}\ }\textbf {\bibinfo {volume} {471}},\ \bibinfo {pages}
  {331--351} (\bibinfo {year} {1996})},\ \Eprint
  {http://arxiv.org/abs/astro-ph/9611094} {arXiv:astro-ph/9611094 [astro-ph]}
  \BibitemShut {NoStop}%
\bibitem [{\citenamefont {Gava}\ \emph {et~al.}(2009)\citenamefont {Gava},
  \citenamefont {Kneller}, \citenamefont {Volpe},\ and\ \citenamefont
  {McLaughlin}}]{Gava:2009pj}%
  \BibitemOpen
  \bibfield  {author} {\bibinfo {author} {\bibfnamefont {Jerome}\ \bibnamefont
  {Gava}}, \bibinfo {author} {\bibfnamefont {James}\ \bibnamefont {Kneller}},
  \bibinfo {author} {\bibfnamefont {Cristina}\ \bibnamefont {Volpe}}, \ and\
  \bibinfo {author} {\bibfnamefont {G.~C.}\ \bibnamefont {McLaughlin}},\
  }\bibfield  {title} {\enquote {\bibinfo {title} {{A Dynamical collective
  calculation of supernova neutrino signals}},}\ }\href {\doibase
  10.1103/PhysRevLett.103.071101} {\bibfield  {journal} {\bibinfo  {journal}
  {Phys. Rev. Lett.}\ }\textbf {\bibinfo {volume} {103}},\ \bibinfo {pages}
  {071101} (\bibinfo {year} {2009})},\ \Eprint {http://arxiv.org/abs/0902.0317}
  {arXiv:0902.0317 [hep-ph]} \BibitemShut {NoStop}%
\bibitem [{\citenamefont {Horiuchi}\ \emph {et~al.}(2009)\citenamefont
  {Horiuchi}, \citenamefont {Beacom},\ and\ \citenamefont
  {Dwek}}]{Horiuchi:2008jz}%
  \BibitemOpen
  \bibfield  {author} {\bibinfo {author} {\bibfnamefont {Shunsaku}\
  \bibnamefont {Horiuchi}}, \bibinfo {author} {\bibfnamefont {John~F.}\
  \bibnamefont {Beacom}}, \ and\ \bibinfo {author} {\bibfnamefont {Eli}\
  \bibnamefont {Dwek}},\ }\bibfield  {title} {\enquote {\bibinfo {title} {{The
  Diffuse Supernova Neutrino Background is detectable in Super-Kamiokande}},}\
  }\href {\doibase 10.1103/PhysRevD.79.083013} {\bibfield  {journal} {\bibinfo
  {journal} {Phys. Rev.}\ }\textbf {\bibinfo {volume} {D79}},\ \bibinfo {pages}
  {083013} (\bibinfo {year} {2009})},\ \Eprint {http://arxiv.org/abs/0812.3157}
  {arXiv:0812.3157 [astro-ph]} \BibitemShut {NoStop}%
\bibitem [{\citenamefont {Beacom}(2010)}]{Beacom:2010kk}%
  \BibitemOpen
  \bibfield  {author} {\bibinfo {author} {\bibfnamefont {John~F.}\ \bibnamefont
  {Beacom}},\ }\bibfield  {title} {\enquote {\bibinfo {title} {{The Diffuse
  Supernova Neutrino Background}},}\ }\href {\doibase
  10.1146/annurev.nucl.010909.083331} {\bibfield  {journal} {\bibinfo
  {journal} {Ann. Rev. Nucl. Part. Sci.}\ }\textbf {\bibinfo {volume} {60}},\
  \bibinfo {pages} {439--462} (\bibinfo {year} {2010})},\ \Eprint
  {http://arxiv.org/abs/1004.3311} {arXiv:1004.3311 [astro-ph.HE]} \BibitemShut
  {NoStop}%
\bibitem [{\citenamefont {Mirizzi}\ \emph {et~al.}(2016)\citenamefont
  {Mirizzi}, \citenamefont {Tamborra}, \citenamefont {Janka}, \citenamefont
  {Saviano}, \citenamefont {Scholberg}, \citenamefont {Bollig}, \citenamefont
  {Hudepohl},\ and\ \citenamefont {Chakraborty}}]{Mirizzi:2015eza}%
  \BibitemOpen
  \bibfield  {author} {\bibinfo {author} {\bibfnamefont {Alessandro}\
  \bibnamefont {Mirizzi}}, \bibinfo {author} {\bibfnamefont {Irene}\
  \bibnamefont {Tamborra}}, \bibinfo {author} {\bibfnamefont {Hans-Thomas}\
  \bibnamefont {Janka}}, \bibinfo {author} {\bibfnamefont {Ninetta}\
  \bibnamefont {Saviano}}, \bibinfo {author} {\bibfnamefont {Kate}\
  \bibnamefont {Scholberg}}, \bibinfo {author} {\bibfnamefont {Robert}\
  \bibnamefont {Bollig}}, \bibinfo {author} {\bibfnamefont {Lorenz}\
  \bibnamefont {Hudepohl}}, \ and\ \bibinfo {author} {\bibfnamefont {Sovan}\
  \bibnamefont {Chakraborty}},\ }\bibfield  {title} {\enquote {\bibinfo {title}
  {{Supernova Neutrinos: Production, Oscillations and Detection}},}\ }\href
  {\doibase 10.1393/ncr/i2016-10120-8} {\bibfield  {journal} {\bibinfo
  {journal} {Riv. Nuovo Cim.}\ }\textbf {\bibinfo {volume} {39}},\ \bibinfo
  {pages} {1--112} (\bibinfo {year} {2016})},\ \Eprint
  {http://arxiv.org/abs/1508.00785} {arXiv:1508.00785 [astro-ph.HE]}
  \BibitemShut {NoStop}%
\bibitem [{\citenamefont {Horiuchi}\ \emph {et~al.}(2018)\citenamefont
  {Horiuchi}, \citenamefont {Sumiyoshi}, \citenamefont {Nakamura},
  \citenamefont {Fischer}, \citenamefont {Summa}, \citenamefont {Takiwaki},
  \citenamefont {Janka},\ and\ \citenamefont {Kotake}}]{Horiuchi:2017qja}%
  \BibitemOpen
  \bibfield  {author} {\bibinfo {author} {\bibfnamefont {Shunsaku}\
  \bibnamefont {Horiuchi}}, \bibinfo {author} {\bibfnamefont {Kohsuke}\
  \bibnamefont {Sumiyoshi}}, \bibinfo {author} {\bibfnamefont {Ko}~\bibnamefont
  {Nakamura}}, \bibinfo {author} {\bibfnamefont {Tobias}\ \bibnamefont
  {Fischer}}, \bibinfo {author} {\bibfnamefont {Alexander}\ \bibnamefont
  {Summa}}, \bibinfo {author} {\bibfnamefont {Tomoya}\ \bibnamefont
  {Takiwaki}}, \bibinfo {author} {\bibfnamefont {Hans-Thomas}\ \bibnamefont
  {Janka}}, \ and\ \bibinfo {author} {\bibfnamefont {Kei}\ \bibnamefont
  {Kotake}},\ }\bibfield  {title} {\enquote {\bibinfo {title} {{Diffuse
  Supernova Neutrino Background from extensive core-collapse simulations of
  $8-100 M_\odot$ progenitors}},}\ }\href {\doibase 10.1093/mnras/stx3271}
  {\bibfield  {journal} {\bibinfo  {journal} {Mon. Not. Roy. Astron. Soc.}\
  }\textbf {\bibinfo {volume} {475}},\ \bibinfo {pages} {1363} (\bibinfo {year}
  {2018})},\ \Eprint {http://arxiv.org/abs/1709.06567} {arXiv:1709.06567
  [astro-ph.HE]} \BibitemShut {NoStop}%
\bibitem [{\citenamefont {Pastor}\ and\ \citenamefont
  {Raffelt}(2002)}]{Pastor:2002we}%
  \BibitemOpen
  \bibfield  {author} {\bibinfo {author} {\bibfnamefont {Sergio}\ \bibnamefont
  {Pastor}}\ and\ \bibinfo {author} {\bibfnamefont {Georg}\ \bibnamefont
  {Raffelt}},\ }\bibfield  {title} {\enquote {\bibinfo {title} {Flavor
  oscillations in the supernova hot bubble region: Nonlinear effects of
  neutrino background},}\ }\href@noop {} {\bibfield  {journal} {\bibinfo
  {journal} {Phys. Rev. Lett.}\ }\textbf {\bibinfo {volume} {89}},\ \bibinfo
  {pages} {191101} (\bibinfo {year} {2002})},\ \Eprint
  {http://arxiv.org/abs/astro-ph/0207281} {astro-ph/0207281} \BibitemShut
  {NoStop}%
\bibitem [{\citenamefont {Duan}\ \emph
  {et~al.}(2006{\natexlab{a}})\citenamefont {Duan}, \citenamefont {Fuller},
  \citenamefont {Carlson},\ and\ \citenamefont {Qian}}]{duan:2006an}%
  \BibitemOpen
  \bibfield  {author} {\bibinfo {author} {\bibfnamefont {Huaiyu}\ \bibnamefont
  {Duan}}, \bibinfo {author} {\bibfnamefont {George~M.}\ \bibnamefont
  {Fuller}}, \bibinfo {author} {\bibfnamefont {J}~\bibnamefont {Carlson}}, \
  and\ \bibinfo {author} {\bibfnamefont {Yong-Zhong}\ \bibnamefont {Qian}},\
  }\bibfield  {title} {\enquote {\bibinfo {title} {{Simulation of Coherent
  Non-Linear Neutrino Flavor Transformation in the Supernova Environment. 1.
  Correlated Neutrino Trajectories}},}\ }\href {\doibase
  10.1103/PhysRevD.74.105014} {\bibfield  {journal} {\bibinfo  {journal} {Phys.
  Rev.}\ }\textbf {\bibinfo {volume} {D74}},\ \bibinfo {pages} {105014}
  (\bibinfo {year} {2006}{\natexlab{a}})},\ \Eprint
  {http://arxiv.org/abs/astro-ph/0606616} {arXiv:astro-ph/0606616 [astro-ph]}
  \BibitemShut {NoStop}%
\bibitem [{\citenamefont {Duan}\ \emph
  {et~al.}(2006{\natexlab{b}})\citenamefont {Duan}, \citenamefont {Fuller},
  \citenamefont {Carlson},\ and\ \citenamefont {Qian}}]{duan:2006jv}%
  \BibitemOpen
  \bibfield  {author} {\bibinfo {author} {\bibfnamefont {Huaiyu}\ \bibnamefont
  {Duan}}, \bibinfo {author} {\bibfnamefont {George~M.}\ \bibnamefont
  {Fuller}}, \bibinfo {author} {\bibfnamefont {J.}~\bibnamefont {Carlson}}, \
  and\ \bibinfo {author} {\bibfnamefont {Yong-Zhong}\ \bibnamefont {Qian}},\
  }\bibfield  {title} {\enquote {\bibinfo {title} {{Coherent Development of
  Neutrino Flavor in the Supernova Environment}},}\ }\href {\doibase
  10.1103/PhysRevLett.97.241101} {\bibfield  {journal} {\bibinfo  {journal}
  {Phys. Rev. Lett.}\ }\textbf {\bibinfo {volume} {97}},\ \bibinfo {pages}
  {241101} (\bibinfo {year} {2006}{\natexlab{b}})},\ \Eprint
  {http://arxiv.org/abs/astro-ph/0608050} {arXiv:astro-ph/0608050 [astro-ph]}
  \BibitemShut {NoStop}%
\bibitem [{\citenamefont {Duan}\ \emph {et~al.}(2010)\citenamefont {Duan},
  \citenamefont {Fuller},\ and\ \citenamefont {Qian}}]{duan:2010bg}%
  \BibitemOpen
  \bibfield  {author} {\bibinfo {author} {\bibfnamefont {Huaiyu}\ \bibnamefont
  {Duan}}, \bibinfo {author} {\bibfnamefont {George~M.}\ \bibnamefont
  {Fuller}}, \ and\ \bibinfo {author} {\bibfnamefont {Yong-Zhong}\ \bibnamefont
  {Qian}},\ }\bibfield  {title} {\enquote {\bibinfo {title} {{Collective
  Neutrino Oscillations}},}\ }\href {\doibase
  10.1146/annurev.nucl.012809.104524} {\bibfield  {journal} {\bibinfo
  {journal} {Ann. Rev. Nucl. Part. Sci.}\ }\textbf {\bibinfo {volume} {60}},\
  \bibinfo {pages} {569--594} (\bibinfo {year} {2010})},\ \Eprint
  {http://arxiv.org/abs/1001.2799} {arXiv:1001.2799 [hep-ph]} \BibitemShut
  {NoStop}%
\bibitem [{\citenamefont {Chakraborty}\ \emph
  {et~al.}(2016{\natexlab{a}})\citenamefont {Chakraborty}, \citenamefont
  {Hansen}, \citenamefont {Izaguirre},\ and\ \citenamefont
  {Raffelt}}]{Chakraborty:2016yeg}%
  \BibitemOpen
  \bibfield  {author} {\bibinfo {author} {\bibfnamefont {Sovan}\ \bibnamefont
  {Chakraborty}}, \bibinfo {author} {\bibfnamefont {Rasmus}\ \bibnamefont
  {Hansen}}, \bibinfo {author} {\bibfnamefont {Ignacio}\ \bibnamefont
  {Izaguirre}}, \ and\ \bibinfo {author} {\bibfnamefont {Georg}\ \bibnamefont
  {Raffelt}},\ }\bibfield  {title} {\enquote {\bibinfo {title} {{Collective
  neutrino flavor conversion: Recent developments}},}\ }\href {\doibase
  10.1016/j.nuclphysb.2016.02.012} {\bibfield  {journal} {\bibinfo  {journal}
  {Nucl. Phys.}\ }\textbf {\bibinfo {volume} {B908}},\ \bibinfo {pages}
  {366--381} (\bibinfo {year} {2016}{\natexlab{a}})},\ \Eprint
  {http://arxiv.org/abs/1602.02766} {arXiv:1602.02766 [hep-ph]} \BibitemShut
  {NoStop}%
\bibitem [{\citenamefont {Wolfenstein}(1978)}]{Wolfenstein:1977ue}%
  \BibitemOpen
  \bibfield  {author} {\bibinfo {author} {\bibfnamefont {L.}~\bibnamefont
  {Wolfenstein}},\ }\bibfield  {title} {\enquote {\bibinfo {title} {{Neutrino
  Oscillations in Matter}},}\ }\href {\doibase 10.1103/PhysRevD.17.2369}
  {\bibfield  {journal} {\bibinfo  {journal} {Phys. Rev.}\ }\textbf {\bibinfo
  {volume} {D17}},\ \bibinfo {pages} {2369--2374} (\bibinfo {year} {1978})},\
  \bibinfo {note} {[,294(1977)]}\BibitemShut {NoStop}%
\bibitem [{\citenamefont {Wolfenstein}(1979)}]{Wolfenstein:1979ni}%
  \BibitemOpen
  \bibfield  {author} {\bibinfo {author} {\bibfnamefont {L.}~\bibnamefont
  {Wolfenstein}},\ }\bibfield  {title} {\enquote {\bibinfo {title} {{Neutrino
  Oscillations and Stellar Collapse}},}\ }\href {\doibase
  10.1103/PhysRevD.20.2634} {\bibfield  {journal} {\bibinfo  {journal} {Phys.
  Rev.}\ }\textbf {\bibinfo {volume} {D20}},\ \bibinfo {pages} {2634--2635}
  (\bibinfo {year} {1979})}\BibitemShut {NoStop}%
\bibitem [{\citenamefont {Mikheyev}\ and\ \citenamefont
  {Smirnov}(1985)}]{Mikheev:1986gs}%
  \BibitemOpen
  \bibfield  {author} {\bibinfo {author} {\bibfnamefont {S.~P.}\ \bibnamefont
  {Mikheyev}}\ and\ \bibinfo {author} {\bibfnamefont {A.~{\relax Yu}.}\
  \bibnamefont {Smirnov}},\ }\bibfield  {title} {\enquote {\bibinfo {title}
  {{Resonance Amplification of Oscillations in Matter and Spectroscopy of Solar
  Neutrinos}},}\ }\href@noop {} {\bibfield  {journal} {\bibinfo  {journal}
  {Sov. J. Nucl. Phys.}\ }\textbf {\bibinfo {volume} {42}},\ \bibinfo {pages}
  {913--917} (\bibinfo {year} {1985})},\ \bibinfo {note}
  {[,305(1986)]}\BibitemShut {NoStop}%
\bibitem [{\citenamefont {Esteban-Pretel}\ \emph {et~al.}(2008)\citenamefont
  {Esteban-Pretel} \emph {et~al.}}]{EstebanPretel:2008ni}%
  \BibitemOpen
  \bibfield  {author} {\bibinfo {author} {\bibfnamefont {A.}~\bibnamefont
  {Esteban-Pretel}} \emph {et~al.},\ }\bibfield  {title} {\enquote {\bibinfo
  {title} {{Role of dense matter in collective supernova neutrino
  transformations}},}\ }\href {\doibase 10.1103/PhysRevD.78.085012} {\bibfield
  {journal} {\bibinfo  {journal} {Phys. Rev.}\ }\textbf {\bibinfo {volume}
  {D78}},\ \bibinfo {pages} {085012} (\bibinfo {year} {2008})},\ \Eprint
  {http://arxiv.org/abs/0807.0659} {arXiv:0807.0659 [astro-ph]} \BibitemShut
  {NoStop}%
\bibitem [{\citenamefont {Sarikas}\ \emph {et~al.}(2012)\citenamefont
  {Sarikas}, \citenamefont {Raffelt}, \citenamefont {Hudepohl},\ and\
  \citenamefont {Janka}}]{Sarikas:2011am}%
  \BibitemOpen
  \bibfield  {author} {\bibinfo {author} {\bibfnamefont {Srdjan}\ \bibnamefont
  {Sarikas}}, \bibinfo {author} {\bibfnamefont {Georg~G.}\ \bibnamefont
  {Raffelt}}, \bibinfo {author} {\bibfnamefont {Lorenz}\ \bibnamefont
  {Hudepohl}}, \ and\ \bibinfo {author} {\bibfnamefont {Hans-Thomas}\
  \bibnamefont {Janka}},\ }\bibfield  {title} {\enquote {\bibinfo {title}
  {{Suppression of Self-Induced Flavor Conversion in the Supernova Accretion
  Phase}},}\ }\href {\doibase 10.1103/PhysRevLett.108.061101} {\bibfield
  {journal} {\bibinfo  {journal} {Phys. Rev. Lett.}\ }\textbf {\bibinfo
  {volume} {108}},\ \bibinfo {pages} {061101} (\bibinfo {year} {2012})},\
  \Eprint {http://arxiv.org/abs/1109.3601} {arXiv:1109.3601 [astro-ph.SR]}
  \BibitemShut {NoStop}%
\bibitem [{\citenamefont {Chakraborty}\ \emph {et~al.}(2011)\citenamefont
  {Chakraborty}, \citenamefont {Fischer}, \citenamefont {Mirizzi},
  \citenamefont {Saviano},\ and\ \citenamefont {Tomas}}]{Chakraborty:2011nf}%
  \BibitemOpen
  \bibfield  {author} {\bibinfo {author} {\bibfnamefont {Sovan}\ \bibnamefont
  {Chakraborty}}, \bibinfo {author} {\bibfnamefont {Tobias}\ \bibnamefont
  {Fischer}}, \bibinfo {author} {\bibfnamefont {Alessandro}\ \bibnamefont
  {Mirizzi}}, \bibinfo {author} {\bibfnamefont {Ninetta}\ \bibnamefont
  {Saviano}}, \ and\ \bibinfo {author} {\bibfnamefont {Ricard}\ \bibnamefont
  {Tomas}},\ }\bibfield  {title} {\enquote {\bibinfo {title} {{No collective
  neutrino flavor conversions during the supernova accretion phase}},}\ }\href
  {\doibase 10.1103/PhysRevLett.107.151101} {\bibfield  {journal} {\bibinfo
  {journal} {Phys. Rev. Lett.}\ }\textbf {\bibinfo {volume} {107}},\ \bibinfo
  {pages} {151101} (\bibinfo {year} {2011})},\ \Eprint
  {http://arxiv.org/abs/1104.4031} {arXiv:1104.4031 [hep-ph]} \BibitemShut
  {NoStop}%
\bibitem [{\citenamefont {Zaizen}\ \emph {et~al.}(2018)\citenamefont {Zaizen},
  \citenamefont {Yoshida}, \citenamefont {Sumiyoshi},\ and\ \citenamefont
  {Umeda}}]{Zaizen:2018wfg}%
  \BibitemOpen
  \bibfield  {author} {\bibinfo {author} {\bibfnamefont {Masamichi}\
  \bibnamefont {Zaizen}}, \bibinfo {author} {\bibfnamefont {Takashi}\
  \bibnamefont {Yoshida}}, \bibinfo {author} {\bibfnamefont {Kohsuke}\
  \bibnamefont {Sumiyoshi}}, \ and\ \bibinfo {author} {\bibfnamefont
  {Hideyuki}\ \bibnamefont {Umeda}},\ }\bibfield  {title} {\enquote {\bibinfo
  {title} {{Collective neutrino oscillations and detectabilities in failed
  supernovae}},}\ }\href {\doibase 10.1103/PhysRevD.98.103020} {\bibfield
  {journal} {\bibinfo  {journal} {Phys. Rev.}\ }\textbf {\bibinfo {volume}
  {D98}},\ \bibinfo {pages} {103020} (\bibinfo {year} {2018})},\ \Eprint
  {http://arxiv.org/abs/1811.03320} {arXiv:1811.03320 [astro-ph.HE]}
  \BibitemShut {NoStop}%
\bibitem [{\citenamefont {Duan}\ and\ \citenamefont
  {Friedland}(2011)}]{Duan:2010bf}%
  \BibitemOpen
  \bibfield  {author} {\bibinfo {author} {\bibfnamefont {Huaiyu}\ \bibnamefont
  {Duan}}\ and\ \bibinfo {author} {\bibfnamefont {Alexander}\ \bibnamefont
  {Friedland}},\ }\bibfield  {title} {\enquote {\bibinfo {title} {{Self-induced
  suppression of collective neutrino oscillations in a supernova}},}\ }\href
  {\doibase 10.1103/PhysRevLett.106.091101} {\bibfield  {journal} {\bibinfo
  {journal} {Phys. Rev. Lett.}\ }\textbf {\bibinfo {volume} {106}},\ \bibinfo
  {pages} {091101} (\bibinfo {year} {2011})},\ \Eprint
  {http://arxiv.org/abs/1006.2359} {arXiv:1006.2359 [hep-ph]} \BibitemShut
  {NoStop}%
\bibitem [{\citenamefont {Duan}\ and\ \citenamefont
  {Shalgar}(2015)}]{Duan:2014gfa}%
  \BibitemOpen
  \bibfield  {author} {\bibinfo {author} {\bibfnamefont {Huaiyu}\ \bibnamefont
  {Duan}}\ and\ \bibinfo {author} {\bibfnamefont {Shashank}\ \bibnamefont
  {Shalgar}},\ }\bibfield  {title} {\enquote {\bibinfo {title} {{Flavor
  instabilities in the neutrino line model}},}\ }\href {\doibase
  10.1016/j.physletb.2015.05.057} {\bibfield  {journal} {\bibinfo  {journal}
  {Phys. Lett.}\ }\textbf {\bibinfo {volume} {B747}},\ \bibinfo {pages}
  {139--143} (\bibinfo {year} {2015})},\ \Eprint
  {http://arxiv.org/abs/1412.7097} {arXiv:1412.7097 [hep-ph]} \BibitemShut
  {NoStop}%
\bibitem [{\citenamefont {Chakraborty}\ \emph
  {et~al.}(2016{\natexlab{b}})\citenamefont {Chakraborty}, \citenamefont
  {Hansen}, \citenamefont {Izaguirre},\ and\ \citenamefont
  {Raffelt}}]{Chakraborty:2015tfa}%
  \BibitemOpen
  \bibfield  {author} {\bibinfo {author} {\bibfnamefont {Sovan}\ \bibnamefont
  {Chakraborty}}, \bibinfo {author} {\bibfnamefont {Rasmus~Sloth}\ \bibnamefont
  {Hansen}}, \bibinfo {author} {\bibfnamefont {Ignacio}\ \bibnamefont
  {Izaguirre}}, \ and\ \bibinfo {author} {\bibfnamefont {Georg}\ \bibnamefont
  {Raffelt}},\ }\bibfield  {title} {\enquote {\bibinfo {title} {{Self-induced
  flavor conversion of supernova neutrinos on small scales}},}\ }\href
  {\doibase 10.1088/1475-7516/2016/01/028} {\bibfield  {journal} {\bibinfo
  {journal} {JCAP}\ }\textbf {\bibinfo {volume} {1601}},\ \bibinfo {pages}
  {028} (\bibinfo {year} {2016}{\natexlab{b}})},\ \Eprint
  {http://arxiv.org/abs/1507.07569} {arXiv:1507.07569 [hep-ph]} \BibitemShut
  {NoStop}%
\bibitem [{\citenamefont {Abbar}\ \emph {et~al.}(2015)\citenamefont {Abbar},
  \citenamefont {Duan},\ and\ \citenamefont {Shalgar}}]{Abbar:2015mca}%
  \BibitemOpen
  \bibfield  {author} {\bibinfo {author} {\bibfnamefont {Sajad}\ \bibnamefont
  {Abbar}}, \bibinfo {author} {\bibfnamefont {Huaiyu}\ \bibnamefont {Duan}}, \
  and\ \bibinfo {author} {\bibfnamefont {Shashank}\ \bibnamefont {Shalgar}},\
  }\bibfield  {title} {\enquote {\bibinfo {title} {{Flavor instabilities in the
  multiangle neutrino line model}},}\ }\href {\doibase
  10.1103/PhysRevD.92.065019} {\bibfield  {journal} {\bibinfo  {journal} {Phys.
  Rev.}\ }\textbf {\bibinfo {volume} {D92}},\ \bibinfo {pages} {065019}
  (\bibinfo {year} {2015})},\ \Eprint {http://arxiv.org/abs/1507.08992}
  {arXiv:1507.08992 [hep-ph]} \BibitemShut {NoStop}%
\bibitem [{\citenamefont {Abbar}\ and\ \citenamefont
  {Duan}(2015)}]{Abbar:2015fwa}%
  \BibitemOpen
  \bibfield  {author} {\bibinfo {author} {\bibfnamefont {Sajad}\ \bibnamefont
  {Abbar}}\ and\ \bibinfo {author} {\bibfnamefont {Huaiyu}\ \bibnamefont
  {Duan}},\ }\bibfield  {title} {\enquote {\bibinfo {title} {{Neutrino flavor
  instabilities in a time-dependent supernova model}},}\ }\href {\doibase
  10.1016/j.physletb.2015.10.019} {\bibfield  {journal} {\bibinfo  {journal}
  {Phys. Lett.}\ }\textbf {\bibinfo {volume} {B751}},\ \bibinfo {pages}
  {43--47} (\bibinfo {year} {2015})},\ \Eprint
  {http://arxiv.org/abs/1509.01538} {arXiv:1509.01538 [astro-ph.HE]}
  \BibitemShut {NoStop}%
\bibitem [{\citenamefont {Dasgupta}\ and\ \citenamefont
  {Mirizzi}(2015)}]{Dasgupta:2015iia}%
  \BibitemOpen
  \bibfield  {author} {\bibinfo {author} {\bibfnamefont {Basudeb}\ \bibnamefont
  {Dasgupta}}\ and\ \bibinfo {author} {\bibfnamefont {Alessandro}\ \bibnamefont
  {Mirizzi}},\ }\bibfield  {title} {\enquote {\bibinfo {title} {{Temporal
  Instability Enables Neutrino Flavor Conversions Deep Inside Supernovae}},}\
  }\href {\doibase 10.1103/PhysRevD.92.125030} {\bibfield  {journal} {\bibinfo
  {journal} {Phys. Rev.}\ }\textbf {\bibinfo {volume} {D92}},\ \bibinfo {pages}
  {125030} (\bibinfo {year} {2015})},\ \Eprint
  {http://arxiv.org/abs/1509.03171} {arXiv:1509.03171 [hep-ph]} \BibitemShut
  {NoStop}%
\bibitem [{\citenamefont {Sawyer}(2005)}]{Sawyer:2005jk}%
  \BibitemOpen
  \bibfield  {author} {\bibinfo {author} {\bibfnamefont {R.~F.}\ \bibnamefont
  {Sawyer}},\ }\bibfield  {title} {\enquote {\bibinfo {title} {{Speed-up of
  neutrino transformations in a supernova environment}},}\ }\href {\doibase
  10.1103/PhysRevD.72.045003} {\bibfield  {journal} {\bibinfo  {journal} {Phys.
  Rev.}\ }\textbf {\bibinfo {volume} {D72}},\ \bibinfo {pages} {045003}
  (\bibinfo {year} {2005})},\ \Eprint {http://arxiv.org/abs/hep-ph/0503013}
  {arXiv:hep-ph/0503013 [hep-ph]} \BibitemShut {NoStop}%
\bibitem [{\citenamefont {Sawyer}(2016)}]{Sawyer:2015dsa}%
  \BibitemOpen
  \bibfield  {author} {\bibinfo {author} {\bibfnamefont {R.~F.}\ \bibnamefont
  {Sawyer}},\ }\bibfield  {title} {\enquote {\bibinfo {title} {{Neutrino cloud
  instabilities just above the neutrino sphere of a supernova}},}\ }\href
  {\doibase 10.1103/PhysRevLett.116.081101} {\bibfield  {journal} {\bibinfo
  {journal} {Phys. Rev. Lett.}\ }\textbf {\bibinfo {volume} {116}},\ \bibinfo
  {pages} {081101} (\bibinfo {year} {2016})},\ \Eprint
  {http://arxiv.org/abs/1509.03323} {arXiv:1509.03323 [astro-ph.HE]}
  \BibitemShut {NoStop}%
\bibitem [{\citenamefont {Chakraborty}\ \emph
  {et~al.}(2016{\natexlab{c}})\citenamefont {Chakraborty}, \citenamefont
  {Hansen}, \citenamefont {Izaguirre},\ and\ \citenamefont
  {Raffelt}}]{Chakraborty:2016lct}%
  \BibitemOpen
  \bibfield  {author} {\bibinfo {author} {\bibfnamefont {Sovan}\ \bibnamefont
  {Chakraborty}}, \bibinfo {author} {\bibfnamefont {Rasmus~Sloth}\ \bibnamefont
  {Hansen}}, \bibinfo {author} {\bibfnamefont {Ignacio}\ \bibnamefont
  {Izaguirre}}, \ and\ \bibinfo {author} {\bibfnamefont {Georg}\ \bibnamefont
  {Raffelt}},\ }\bibfield  {title} {\enquote {\bibinfo {title} {{Self-induced
  neutrino flavor conversion without flavor mixing}},}\ }\href {\doibase
  10.1088/1475-7516/2016/03/042} {\bibfield  {journal} {\bibinfo  {journal}
  {JCAP}\ }\textbf {\bibinfo {volume} {1603}},\ \bibinfo {pages} {042}
  (\bibinfo {year} {2016}{\natexlab{c}})},\ \Eprint
  {http://arxiv.org/abs/1602.00698} {arXiv:1602.00698 [hep-ph]} \BibitemShut
  {NoStop}%
\bibitem [{\citenamefont {Izaguirre}\ \emph {et~al.}(2017)\citenamefont
  {Izaguirre}, \citenamefont {Raffelt},\ and\ \citenamefont
  {Tamborra}}]{Izaguirre:2016gsx}%
  \BibitemOpen
  \bibfield  {author} {\bibinfo {author} {\bibfnamefont {Ignacio}\ \bibnamefont
  {Izaguirre}}, \bibinfo {author} {\bibfnamefont {Georg}\ \bibnamefont
  {Raffelt}}, \ and\ \bibinfo {author} {\bibfnamefont {Irene}\ \bibnamefont
  {Tamborra}},\ }\bibfield  {title} {\enquote {\bibinfo {title} {{Fast Pairwise
  Conversion of Supernova Neutrinos: A Dispersion-Relation Approach}},}\ }\href
  {\doibase 10.1103/PhysRevLett.118.021101} {\bibfield  {journal} {\bibinfo
  {journal} {Phys. Rev. Lett.}\ }\textbf {\bibinfo {volume} {118}},\ \bibinfo
  {pages} {021101} (\bibinfo {year} {2017})},\ \Eprint
  {http://arxiv.org/abs/1610.01612} {arXiv:1610.01612 [hep-ph]} \BibitemShut
  {NoStop}%
\bibitem [{\citenamefont {Wu}\ and\ \citenamefont
  {Tamborra}(2017)}]{Wu:2017qpc}%
  \BibitemOpen
  \bibfield  {author} {\bibinfo {author} {\bibfnamefont {Meng-Ru}\ \bibnamefont
  {Wu}}\ and\ \bibinfo {author} {\bibfnamefont {Irene}\ \bibnamefont
  {Tamborra}},\ }\bibfield  {title} {\enquote {\bibinfo {title} {{Fast neutrino
  conversions: Ubiquitous in compact binary merger remnants}},}\ }\href
  {\doibase 10.1103/PhysRevD.95.103007} {\bibfield  {journal} {\bibinfo
  {journal} {Phys. Rev.}\ }\textbf {\bibinfo {volume} {D95}},\ \bibinfo {pages}
  {103007} (\bibinfo {year} {2017})},\ \Eprint
  {http://arxiv.org/abs/1701.06580} {arXiv:1701.06580 [astro-ph.HE]}
  \BibitemShut {NoStop}%
\bibitem [{\citenamefont {Capozzi}\ \emph {et~al.}(2017)\citenamefont
  {Capozzi}, \citenamefont {Dasgupta}, \citenamefont {Lisi}, \citenamefont
  {Marrone},\ and\ \citenamefont {Mirizzi}}]{Capozzi:2017gqd}%
  \BibitemOpen
  \bibfield  {author} {\bibinfo {author} {\bibfnamefont {Francesco}\
  \bibnamefont {Capozzi}}, \bibinfo {author} {\bibfnamefont {Basudeb}\
  \bibnamefont {Dasgupta}}, \bibinfo {author} {\bibfnamefont {Eligio}\
  \bibnamefont {Lisi}}, \bibinfo {author} {\bibfnamefont {Antonio}\
  \bibnamefont {Marrone}}, \ and\ \bibinfo {author} {\bibfnamefont
  {Alessandro}\ \bibnamefont {Mirizzi}},\ }\bibfield  {title} {\enquote
  {\bibinfo {title} {{Fast flavor conversions of supernova neutrinos:
  Classifying instabilities via dispersion relations}},}\ }\href {\doibase
  10.1103/PhysRevD.96.043016} {\bibfield  {journal} {\bibinfo  {journal} {Phys.
  Rev.}\ }\textbf {\bibinfo {volume} {D96}},\ \bibinfo {pages} {043016}
  (\bibinfo {year} {2017})},\ \Eprint {http://arxiv.org/abs/1706.03360}
  {arXiv:1706.03360 [hep-ph]} \BibitemShut {NoStop}%
\bibitem [{\citenamefont {Dasgupta}\ \emph {et~al.}(2017)\citenamefont
  {Dasgupta}, \citenamefont {Mirizzi},\ and\ \citenamefont
  {Sen}}]{Dasgupta:2016dbv}%
  \BibitemOpen
  \bibfield  {author} {\bibinfo {author} {\bibfnamefont {Basudeb}\ \bibnamefont
  {Dasgupta}}, \bibinfo {author} {\bibfnamefont {Alessandro}\ \bibnamefont
  {Mirizzi}}, \ and\ \bibinfo {author} {\bibfnamefont {Manibrata}\ \bibnamefont
  {Sen}},\ }\bibfield  {title} {\enquote {\bibinfo {title} {{Fast neutrino
  flavor conversions near the supernova core with realistic flavor-dependent
  angular distributions}},}\ }\href {\doibase 10.1088/1475-7516/2017/02/019}
  {\bibfield  {journal} {\bibinfo  {journal} {JCAP}\ }\textbf {\bibinfo
  {volume} {1702}},\ \bibinfo {pages} {019} (\bibinfo {year} {2017})},\ \Eprint
  {http://arxiv.org/abs/1609.00528} {arXiv:1609.00528 [hep-ph]} \BibitemShut
  {NoStop}%
\bibitem [{\citenamefont {Abbar}\ and\ \citenamefont
  {Duan}(2018)}]{Abbar:2017pkh}%
  \BibitemOpen
  \bibfield  {author} {\bibinfo {author} {\bibfnamefont {Sajad}\ \bibnamefont
  {Abbar}}\ and\ \bibinfo {author} {\bibfnamefont {Huaiyu}\ \bibnamefont
  {Duan}},\ }\bibfield  {title} {\enquote {\bibinfo {title} {{Fast neutrino
  flavor conversion: roles of dense matter and spectrum crossing}},}\ }\href
  {\doibase 10.1103/PhysRevD.98.043014} {\bibfield  {journal} {\bibinfo
  {journal} {Phys. Rev.}\ }\textbf {\bibinfo {volume} {D98}},\ \bibinfo {pages}
  {043014} (\bibinfo {year} {2018})},\ \Eprint
  {http://arxiv.org/abs/1712.07013} {arXiv:1712.07013 [hep-ph]} \BibitemShut
  {NoStop}%
\bibitem [{\citenamefont {Abbar}\ and\ \citenamefont
  {Volpe}(2019)}]{Abbar:2018beu}%
  \BibitemOpen
  \bibfield  {author} {\bibinfo {author} {\bibfnamefont {Sajad}\ \bibnamefont
  {Abbar}}\ and\ \bibinfo {author} {\bibfnamefont {Maria~Cristina}\
  \bibnamefont {Volpe}},\ }\bibfield  {title} {\enquote {\bibinfo {title} {{On
  Fast Neutrino Flavor Conversion Modes in the Nonlinear Regime}},}\ }\href
  {\doibase 10.1016/j.physletb.2019.02.002} {\bibfield  {journal} {\bibinfo
  {journal} {Phys. Lett.}\ }\textbf {\bibinfo {volume} {B790}},\ \bibinfo
  {pages} {545--550} (\bibinfo {year} {2019})},\ \Eprint
  {http://arxiv.org/abs/1811.04215} {arXiv:1811.04215 [astro-ph.HE]}
  \BibitemShut {NoStop}%
\bibitem [{\citenamefont {Capozzi}\ \emph {et~al.}(2019)\citenamefont
  {Capozzi}, \citenamefont {Dasgupta}, \citenamefont {Mirizzi}, \citenamefont
  {Sen},\ and\ \citenamefont {Sigl}}]{Capozzi:2018clo}%
  \BibitemOpen
  \bibfield  {author} {\bibinfo {author} {\bibfnamefont {Francesco}\
  \bibnamefont {Capozzi}}, \bibinfo {author} {\bibfnamefont {Basudeb}\
  \bibnamefont {Dasgupta}}, \bibinfo {author} {\bibfnamefont {Alessandro}\
  \bibnamefont {Mirizzi}}, \bibinfo {author} {\bibfnamefont {Manibrata}\
  \bibnamefont {Sen}}, \ and\ \bibinfo {author} {\bibfnamefont {Günter}\
  \bibnamefont {Sigl}},\ }\bibfield  {title} {\enquote {\bibinfo {title}
  {{Collisional triggering of fast flavor conversions of supernova
  neutrinos}},}\ }\href {\doibase 10.1103/PhysRevLett.122.091101} {\bibfield
  {journal} {\bibinfo  {journal} {Phys. Rev. Lett.}\ }\textbf {\bibinfo
  {volume} {122}},\ \bibinfo {pages} {091101} (\bibinfo {year} {2019})},\
  \Eprint {http://arxiv.org/abs/1808.06618} {arXiv:1808.06618 [hep-ph]}
  \BibitemShut {NoStop}%
\bibitem [{\citenamefont {Tamborra}\ \emph {et~al.}(2017)\citenamefont
  {Tamborra}, \citenamefont {Huedepohl}, \citenamefont {Raffelt},\ and\
  \citenamefont {Janka}}]{Tamborra:2017ubu}%
  \BibitemOpen
  \bibfield  {author} {\bibinfo {author} {\bibfnamefont {Irene}\ \bibnamefont
  {Tamborra}}, \bibinfo {author} {\bibfnamefont {Lorenz}\ \bibnamefont
  {Huedepohl}}, \bibinfo {author} {\bibfnamefont {Georg}\ \bibnamefont
  {Raffelt}}, \ and\ \bibinfo {author} {\bibfnamefont {Hans-Thomas}\
  \bibnamefont {Janka}},\ }\bibfield  {title} {\enquote {\bibinfo {title}
  {{Flavor-dependent neutrino angular distribution in core-collapse
  supernovae}},}\ }\href {\doibase 10.3847/1538-4357/aa6a18} {\bibfield
  {journal} {\bibinfo  {journal} {Astrophys. J.}\ }\textbf {\bibinfo {volume}
  {839}},\ \bibinfo {pages} {132} (\bibinfo {year} {2017})},\ \Eprint
  {http://arxiv.org/abs/1702.00060} {arXiv:1702.00060 [astro-ph.HE]}
  \BibitemShut {NoStop}%
\bibitem [{\citenamefont {Tamborra}\ \emph {et~al.}(2014)\citenamefont
  {Tamborra}, \citenamefont {Hanke}, \citenamefont {Janka}, \citenamefont
  {M{\"u}ller}, \citenamefont {Raffelt},\ and\ \citenamefont
  {Marek}}]{Tamborra:2014aua}%
  \BibitemOpen
  \bibfield  {author} {\bibinfo {author} {\bibfnamefont {Irene}\ \bibnamefont
  {Tamborra}}, \bibinfo {author} {\bibfnamefont {Florian}\ \bibnamefont
  {Hanke}}, \bibinfo {author} {\bibfnamefont {Hans-Thomas}\ \bibnamefont
  {Janka}}, \bibinfo {author} {\bibfnamefont {Bernhard}\ \bibnamefont
  {M{\"u}ller}}, \bibinfo {author} {\bibfnamefont {Georg~G.}\ \bibnamefont
  {Raffelt}}, \ and\ \bibinfo {author} {\bibfnamefont {Andreas}\ \bibnamefont
  {Marek}},\ }\bibfield  {title} {\enquote {\bibinfo {title} {{Self-sustained
  asymmetry of lepton-number emission: A new phenomenon during the supernova
  shock-accretion phase in three dimensions}},}\ }\href {\doibase
  10.1088/0004-637X/792/2/96} {\bibfield  {journal} {\bibinfo  {journal}
  {Astrophys. J.}\ }\textbf {\bibinfo {volume} {792}},\ \bibinfo {pages} {96}
  (\bibinfo {year} {2014})},\ \Eprint {http://arxiv.org/abs/1402.5418}
  {arXiv:1402.5418 [astro-ph.SR]} \BibitemShut {NoStop}%
\bibitem [{\citenamefont {Sumiyoshi}\ and\ \citenamefont
  {Yamada}(2012)}]{Sumiyoshi:2012za}%
  \BibitemOpen
  \bibfield  {author} {\bibinfo {author} {\bibfnamefont {Kohsuke}\ \bibnamefont
  {Sumiyoshi}}\ and\ \bibinfo {author} {\bibfnamefont {Shoichi}\ \bibnamefont
  {Yamada}},\ }\bibfield  {title} {\enquote {\bibinfo {title} {{Neutrino
  Transfer in Three Dimensions for Core-Collapse Supernovae. I. Static
  Configurations}},}\ }\href {\doibase 10.1088/0067-0049/199/1/17} {\bibfield
  {journal} {\bibinfo  {journal} {Astrophys. J. Suppl.}\ }\textbf {\bibinfo
  {volume} {199}},\ \bibinfo {pages} {17} (\bibinfo {year} {2012})},\ \Eprint
  {http://arxiv.org/abs/1201.2244} {arXiv:1201.2244 [astro-ph.HE]} \BibitemShut
  {NoStop}%
\bibitem [{\citenamefont {Sumiyoshi}\ \emph {et~al.}(2015)\citenamefont
  {Sumiyoshi}, \citenamefont {Takiwaki}, \citenamefont {Matsufuru},\ and\
  \citenamefont {Yamada}}]{Sumiyoshi:2014qua}%
  \BibitemOpen
  \bibfield  {author} {\bibinfo {author} {\bibfnamefont {K.}~\bibnamefont
  {Sumiyoshi}}, \bibinfo {author} {\bibfnamefont {T.}~\bibnamefont {Takiwaki}},
  \bibinfo {author} {\bibfnamefont {H.}~\bibnamefont {Matsufuru}}, \ and\
  \bibinfo {author} {\bibfnamefont {S.}~\bibnamefont {Yamada}},\ }\bibfield
  {title} {\enquote {\bibinfo {title} {{Multi-dimensional Features of Neutrino
  Transfer in Core-Collapse Supernovae}},}\ }\href {\doibase
  10.1088/0067-0049/216/1/5} {\bibfield  {journal} {\bibinfo  {journal}
  {Astrophys. J. Suppl.}\ }\textbf {\bibinfo {volume} {216}},\ \bibinfo {pages}
  {5} (\bibinfo {year} {2015})},\ \Eprint {http://arxiv.org/abs/1403.4476}
  {arXiv:1403.4476 [astro-ph.HE]} \BibitemShut {NoStop}%
\bibitem [{\citenamefont {Nagakura}\ \emph {et~al.}(2018)\citenamefont
  {Nagakura}, \citenamefont {Iwakami}, \citenamefont {Furusawa}, \citenamefont
  {Okawa}, \citenamefont {Harada}, \citenamefont {Sumiyoshi}, \citenamefont
  {Yamada}, \citenamefont {Matsufuru},\ and\ \citenamefont
  {Imakura}}]{Nagakura:2017mnp}%
  \BibitemOpen
  \bibfield  {author} {\bibinfo {author} {\bibfnamefont {Hiroki}\ \bibnamefont
  {Nagakura}}, \bibinfo {author} {\bibfnamefont {Wakana}\ \bibnamefont
  {Iwakami}}, \bibinfo {author} {\bibfnamefont {Shun}\ \bibnamefont
  {Furusawa}}, \bibinfo {author} {\bibfnamefont {Hirotada}\ \bibnamefont
  {Okawa}}, \bibinfo {author} {\bibfnamefont {Akira}\ \bibnamefont {Harada}},
  \bibinfo {author} {\bibfnamefont {Kohsuke}\ \bibnamefont {Sumiyoshi}},
  \bibinfo {author} {\bibfnamefont {Shoichi}\ \bibnamefont {Yamada}}, \bibinfo
  {author} {\bibfnamefont {Hideo}\ \bibnamefont {Matsufuru}}, \ and\ \bibinfo
  {author} {\bibfnamefont {Akira}\ \bibnamefont {Imakura}},\ }\bibfield
  {title} {\enquote {\bibinfo {title} {{Simulations of core-collapse supernovae
  in spatial axisymmetry with full Boltzmann neutrino transport}},}\ }\href
  {\doibase 10.3847/1538-4357/aaac29} {\bibfield  {journal} {\bibinfo
  {journal} {Astrophys. J.}\ }\textbf {\bibinfo {volume} {854}},\ \bibinfo
  {pages} {136} (\bibinfo {year} {2018})},\ \Eprint
  {http://arxiv.org/abs/1702.01752} {arXiv:1702.01752 [astro-ph.HE]}
  \BibitemShut {NoStop}%
\bibitem [{\citenamefont {Sigl}\ and\ \citenamefont
  {Raffelt}(1993)}]{Sigl:1992fn}%
  \BibitemOpen
  \bibfield  {author} {\bibinfo {author} {\bibfnamefont {G.}~\bibnamefont
  {Sigl}}\ and\ \bibinfo {author} {\bibfnamefont {G.}~\bibnamefont {Raffelt}},\
  }\bibfield  {title} {\enquote {\bibinfo {title} {{General kinetic description
  of relativistic mixed neutrinos}},}\ }\href {\doibase
  10.1016/0550-3213(93)90175-O} {\bibfield  {journal} {\bibinfo  {journal}
  {Nucl. Phys.}\ }\textbf {\bibinfo {volume} {B406}},\ \bibinfo {pages}
  {423--451} (\bibinfo {year} {1993})}\BibitemShut {NoStop}%
\bibitem [{\citenamefont {Banerjee}\ \emph {et~al.}(2011)\citenamefont
  {Banerjee}, \citenamefont {Dighe},\ and\ \citenamefont
  {Raffelt}}]{Banerjee:2011fj}%
  \BibitemOpen
  \bibfield  {author} {\bibinfo {author} {\bibfnamefont {Arka}\ \bibnamefont
  {Banerjee}}, \bibinfo {author} {\bibfnamefont {Amol}\ \bibnamefont {Dighe}},
  \ and\ \bibinfo {author} {\bibfnamefont {Georg}\ \bibnamefont {Raffelt}},\
  }\bibfield  {title} {\enquote {\bibinfo {title} {{Linearized flavor-stability
  analysis of dense neutrino streams}},}\ }\href {\doibase
  10.1103/PhysRevD.84.053013} {\bibfield  {journal} {\bibinfo  {journal} {Phys.
  Rev.}\ }\textbf {\bibinfo {volume} {D84}},\ \bibinfo {pages} {053013}
  (\bibinfo {year} {2011})},\ \Eprint {http://arxiv.org/abs/1107.2308}
  {arXiv:1107.2308 [hep-ph]} \BibitemShut {NoStop}%
\bibitem [{\citenamefont {Strack}\ and\ \citenamefont
  {Burrows}(2005)}]{Strack:2005ux}%
  \BibitemOpen
  \bibfield  {author} {\bibinfo {author} {\bibfnamefont {P.}~\bibnamefont
  {Strack}}\ and\ \bibinfo {author} {\bibfnamefont {A.}~\bibnamefont
  {Burrows}},\ }\bibfield  {title} {\enquote {\bibinfo {title} {A generalized
  boltzmann formalism for oscillating neutrinos},}\ }\href@noop {} {\bibfield
  {journal} {\bibinfo  {journal} {Phys. Rev.}\ }\textbf {\bibinfo {volume}
  {D71}},\ \bibinfo {pages} {093004} (\bibinfo {year} {2005})},\ \Eprint
  {http://arxiv.org/abs/hep-ph/0504035} {hep-ph/0504035} \BibitemShut {NoStop}%
\bibitem [{\citenamefont {Cardall}(2008)}]{Cardall:2007zw}%
  \BibitemOpen
  \bibfield  {author} {\bibinfo {author} {\bibfnamefont {Christian~Y.}\
  \bibnamefont {Cardall}},\ }\bibfield  {title} {\enquote {\bibinfo {title}
  {{Liouville equations for neutrino distribution matrices}},}\ }\href
  {\doibase 10.1103/PhysRevD.78.085017} {\bibfield  {journal} {\bibinfo
  {journal} {Phys. Rev.}\ }\textbf {\bibinfo {volume} {D78}},\ \bibinfo {pages}
  {085017} (\bibinfo {year} {2008})},\ \Eprint {http://arxiv.org/abs/0712.1188}
  {arXiv:0712.1188 [astro-ph]} \BibitemShut {NoStop}%
\bibitem [{\citenamefont {Volpe}\ \emph {et~al.}(2013)\citenamefont {Volpe},
  \citenamefont {Väänänen},\ and\ \citenamefont {Espinoza}}]{Volpe:2013jgr}%
  \BibitemOpen
  \bibfield  {author} {\bibinfo {author} {\bibfnamefont {Cristina}\
  \bibnamefont {Volpe}}, \bibinfo {author} {\bibfnamefont {Daavid}\
  \bibnamefont {Väänänen}}, \ and\ \bibinfo {author} {\bibfnamefont
  {Catalina}\ \bibnamefont {Espinoza}},\ }\bibfield  {title} {\enquote
  {\bibinfo {title} {{Extended evolution equations for neutrino propagation in
  astrophysical and cosmological environments}},}\ }\href {\doibase
  10.1103/PhysRevD.87.113010} {\bibfield  {journal} {\bibinfo  {journal} {Phys.
  Rev.}\ }\textbf {\bibinfo {volume} {D87}},\ \bibinfo {pages} {113010}
  (\bibinfo {year} {2013})},\ \Eprint {http://arxiv.org/abs/1302.2374}
  {arXiv:1302.2374 [hep-ph]} \BibitemShut {NoStop}%
\bibitem [{\citenamefont {Vlasenko}\ \emph {et~al.}(2014)\citenamefont
  {Vlasenko}, \citenamefont {Fuller},\ and\ \citenamefont
  {Cirigliano}}]{Vlasenko:2013fja}%
  \BibitemOpen
  \bibfield  {author} {\bibinfo {author} {\bibfnamefont {Alexey}\ \bibnamefont
  {Vlasenko}}, \bibinfo {author} {\bibfnamefont {George~M.}\ \bibnamefont
  {Fuller}}, \ and\ \bibinfo {author} {\bibfnamefont {Vincenzo}\ \bibnamefont
  {Cirigliano}},\ }\bibfield  {title} {\enquote {\bibinfo {title} {{Neutrino
  Quantum Kinetics}},}\ }\href {\doibase 10.1103/PhysRevD.89.105004} {\bibfield
   {journal} {\bibinfo  {journal} {Phys. Rev.}\ }\textbf {\bibinfo {volume}
  {D89}},\ \bibinfo {pages} {105004} (\bibinfo {year} {2014})},\ \Eprint
  {http://arxiv.org/abs/1309.2628} {arXiv:1309.2628 [hep-ph]} \BibitemShut
  {NoStop}%
\bibitem [{\citenamefont {Fuller}\ \emph {et~al.}(1987)\citenamefont {Fuller},
  \citenamefont {Mayle}, \citenamefont {Wilson},\ and\ \citenamefont
  {Schramm}}]{Fuller:1987aa}%
  \BibitemOpen
  \bibfield  {author} {\bibinfo {author} {\bibfnamefont {George~M.}\
  \bibnamefont {Fuller}}, \bibinfo {author} {\bibfnamefont {Ron~W.}\
  \bibnamefont {Mayle}}, \bibinfo {author} {\bibfnamefont {James~R.}\
  \bibnamefont {Wilson}}, \ and\ \bibinfo {author} {\bibfnamefont {David~N.}\
  \bibnamefont {Schramm}},\ }\bibfield  {title} {\enquote {\bibinfo {title}
  {Resonant neutrino oscillations and stellar collapse},}\ }\href@noop {}
  {\bibfield  {journal} {\bibinfo  {journal} {Astrophys. J.}\ }\textbf
  {\bibinfo {volume} {322}},\ \bibinfo {pages} {795} (\bibinfo {year}
  {1987})}\BibitemShut {NoStop}%
\bibitem [{\citenamefont {N\"{o}tzold}\ and\ \citenamefont
  {Raffelt}(1988)}]{Notzold:1988kx}%
  \BibitemOpen
  \bibfield  {author} {\bibinfo {author} {\bibfnamefont {D.}~\bibnamefont
  {N\"{o}tzold}}\ and\ \bibinfo {author} {\bibfnamefont {G.}~\bibnamefont
  {Raffelt}},\ }\bibfield  {title} {\enquote {\bibinfo {title} {Neutrono
  dispersion at finite temperature and density},}\ }\href@noop {} {\bibfield
  {journal} {\bibinfo  {journal} {Nucl. Phys.}\ }\textbf {\bibinfo {volume}
  {B307}},\ \bibinfo {pages} {924} (\bibinfo {year} {1988})}\BibitemShut
  {NoStop}%
\bibitem [{\citenamefont {Pantaleone}(1992)}]{Pantaleone:1992xh}%
  \BibitemOpen
  \bibfield  {author} {\bibinfo {author} {\bibfnamefont {James~T.}\
  \bibnamefont {Pantaleone}},\ }\bibfield  {title} {\enquote {\bibinfo {title}
  {{Dirac neutrinos in dense matter}},}\ }\href {\doibase
  10.1103/PhysRevD.46.510} {\bibfield  {journal} {\bibinfo  {journal} {Phys.
  Rev.}\ }\textbf {\bibinfo {volume} {D46}},\ \bibinfo {pages} {510--523}
  (\bibinfo {year} {1992})}\BibitemShut {NoStop}%
\bibitem [{\citenamefont {Väänänen}\ and\ \citenamefont
  {Volpe}(2013)}]{Vaananen:2013qja}%
  \BibitemOpen
  \bibfield  {author} {\bibinfo {author} {\bibfnamefont {D.}~\bibnamefont
  {Väänänen}}\ and\ \bibinfo {author} {\bibfnamefont {C.}~\bibnamefont
  {Volpe}},\ }\bibfield  {title} {\enquote {\bibinfo {title} {{Linearizing
  neutrino evolution equations including neutrino-antineutrino pairing
  correlations}},}\ }\href {\doibase 10.1103/PhysRevD.88.065003} {\bibfield
  {journal} {\bibinfo  {journal} {Phys. Rev.}\ }\textbf {\bibinfo {volume}
  {D88}},\ \bibinfo {pages} {065003} (\bibinfo {year} {2013})},\ \Eprint
  {http://arxiv.org/abs/1306.6372} {arXiv:1306.6372 [hep-ph]} \BibitemShut
  {NoStop}%
\bibitem [{\citenamefont {Lattimer}\ and\ \citenamefont
  {Swesty}(1991)}]{Lattimer:1991nc}%
  \BibitemOpen
  \bibfield  {author} {\bibinfo {author} {\bibfnamefont {James~M.}\
  \bibnamefont {Lattimer}}\ and\ \bibinfo {author} {\bibfnamefont {F.~Douglas}\
  \bibnamefont {Swesty}},\ }\bibfield  {title} {\enquote {\bibinfo {title} {{A
  Generalized equation of state for hot, dense matter}},}\ }\href {\doibase
  10.1016/0375-9474(91)90452-C} {\bibfield  {journal} {\bibinfo  {journal}
  {Nucl. Phys.}\ }\textbf {\bibinfo {volume} {A535}},\ \bibinfo {pages}
  {331--376} (\bibinfo {year} {1991})}\BibitemShut {NoStop}%
\bibitem [{\citenamefont {Takiwaki}\ \emph {et~al.}(2012)\citenamefont
  {Takiwaki}, \citenamefont {Kotake},\ and\ \citenamefont
  {Suwa}}]{Takiwaki:2011db}%
  \BibitemOpen
  \bibfield  {author} {\bibinfo {author} {\bibfnamefont {Tomoya}\ \bibnamefont
  {Takiwaki}}, \bibinfo {author} {\bibfnamefont {Kei}\ \bibnamefont {Kotake}},
  \ and\ \bibinfo {author} {\bibfnamefont {Yudai}\ \bibnamefont {Suwa}},\
  }\bibfield  {title} {\enquote {\bibinfo {title} {{Three-dimensional
  Hydrodynamic Core-Collapse Supernova Simulations for an $11.2 M_{\odot}$ Star
  with Spectral Neutrino Transport}},}\ }\href {\doibase
  10.1088/0004-637X/749/2/98} {\bibfield  {journal} {\bibinfo  {journal}
  {Astrophys. J.}\ }\textbf {\bibinfo {volume} {749}},\ \bibinfo {pages} {98}
  (\bibinfo {year} {2012})},\ \Eprint {http://arxiv.org/abs/1108.3989}
  {arXiv:1108.3989 [astro-ph.HE]} \BibitemShut {NoStop}%
\bibitem [{\citenamefont {Takiwaki}\ \emph {et~al.}(2014)\citenamefont
  {Takiwaki}, \citenamefont {Kotake},\ and\ \citenamefont
  {Suwa}}]{Takiwaki:2013cqa}%
  \BibitemOpen
  \bibfield  {author} {\bibinfo {author} {\bibfnamefont {Tomoya}\ \bibnamefont
  {Takiwaki}}, \bibinfo {author} {\bibfnamefont {Kei}\ \bibnamefont {Kotake}},
  \ and\ \bibinfo {author} {\bibfnamefont {Yudai}\ \bibnamefont {Suwa}},\
  }\bibfield  {title} {\enquote {\bibinfo {title} {{A Comparison of Two- and
  Three-dimensional Neutrino-hydrodynamics simulations of Core-collapse
  Supernovae}},}\ }\href {\doibase 10.1088/0004-637X/786/2/83} {\bibfield
  {journal} {\bibinfo  {journal} {Astrophys. J.}\ }\textbf {\bibinfo {volume}
  {786}},\ \bibinfo {pages} {83} (\bibinfo {year} {2014})},\ \Eprint
  {http://arxiv.org/abs/1308.5755} {arXiv:1308.5755 [astro-ph.SR]} \BibitemShut
  {NoStop}%
\bibitem [{\citenamefont {Vartanyan}\ \emph {et~al.}(2019)\citenamefont
  {Vartanyan}, \citenamefont {Burrows}, \citenamefont {Radice}, \citenamefont
  {Skinner},\ and\ \citenamefont {Dolence}}]{Vartanyan:2018iah}%
  \BibitemOpen
  \bibfield  {author} {\bibinfo {author} {\bibfnamefont {David}\ \bibnamefont
  {Vartanyan}}, \bibinfo {author} {\bibfnamefont {Adam}\ \bibnamefont
  {Burrows}}, \bibinfo {author} {\bibfnamefont {David}\ \bibnamefont {Radice}},
  \bibinfo {author} {\bibfnamefont {Aaron}\ \bibnamefont {Skinner}}, \ and\
  \bibinfo {author} {\bibfnamefont {Joshua}\ \bibnamefont {Dolence}},\
  }\bibfield  {title} {\enquote {\bibinfo {title} {{A Successful 3D
  Core-Collapse Supernova Explosion Model}},}\ }\href {\doibase
  10.1093/mnras/sty2585} {\bibfield  {journal} {\bibinfo  {journal} {Mon. Not.
  Roy. Astron. Soc.}\ }\textbf {\bibinfo {volume} {482}},\ \bibinfo {pages}
  {351} (\bibinfo {year} {2019})},\ \Eprint {http://arxiv.org/abs/1809.05106}
  {arXiv:1809.05106 [astro-ph.HE]} \BibitemShut {NoStop}%
\end{thebibliography}%


\end{document}